\documentclass[]{aa}  

\usepackage{graphicx}
\usepackage{txfonts}
\usepackage{multirow}
\usepackage{gensymb}
\usepackage[]{hyperref}
\hypersetup{breaklinks=true,colorlinks=true,urlcolor=blue, linkcolor=blue, citecolor=blue}

\begin{document}

   \title{ALMA chemical survey of disk-outflow sources in Taurus (ALMA-DOT)}

   \subtitle{I. CO, CS, CN, and H$_2$CO around DG Tau B}

   \author{A.\,Garufi \inst{\ref{Firenze}}
   \and L.\,Podio \inst{\ref{Firenze}}
   \and C.\,Codella\inst{\ref{Firenze}, \ref{IPAG}}
   \and K.\,Rygl \inst{\ref{IRA}}
   \and F.\,Bacciotti \inst{\ref{Firenze}}
   \and S.\,Facchini \inst{\ref{ESO_Germany}}
   \and \\ D.\,Fedele \inst{\ref{Firenze}}
   \and A.\,Miotello \inst{\ref{ESO_Germany}}
   \and R.\,Teague \inst{\ref{Michigan}, \ref{CfA}}
   \and L.\,Testi\inst{\ref{ESO_Germany}, \ref{Firenze}}
     }

\institute{INAF, Osservatorio Astrofisico di Arcetri, Largo Enrico Fermi 5, I-50125 Firenze, Italy. \label{Firenze}
  \email{agarufi@arcetri.astro.it}  
     \and Univ. Grenoble Alpes, CNRS, IPAG, F-38000 Grenoble, France \label{IPAG}
     \and INAF$-$Istituto di Radioastronomia \& Italian ALMA Regional Centre, via P.\,Gobetti 101, 40129 Bologna, Italy \label{IRA}
       \and European Southern Observatory, Karl-Schwarzschild-Strasse 2, D-85748 Garching, Germany \label{ESO_Germany}
         \and University of Michigan, Department of Astronomy, 1085 S.\ University, Ann Arbor, MI 48109 \label{Michigan}
         \and Center for Astrophysics | Harvard \& Smithsonian, 60 Garden Street, Cambridge, MA 02138, United States of America \label{CfA}
             }

   \date{Received -; accepted -}

 
 \abstract{The chemical composition of planets is determined by  the distribution of the various molecular species in the protoplanetary disk at the time of their formation. To date, only a handful of disks have been imaged in multiple spectral lines with high spatial resolution. As part of a small campaign devoted to the chemical characterization of disk-outflow sources in Taurus, we report on new ALMA Band 6 ($\sim$1.3 mm) observations with $\sim$0.15\arcsec\ (20 au) resolution toward the embedded young star DG Tau B. Images of the continuum emission reveals a dust disk with rings and, putatively, a leading spiral arm. The disk, as well as the prominent outflow cavities, are detected in CO, H$_2$CO, CS, and CN; instead,  they remain undetected in SO$_2$, HDO, and CH$_3$OH. From the absorption of the back-side outflow, we inferred that the disk emission is optically thick in the inner 50 au. This morphology explains why no line emission is detected from this inner region and poses some limitations toward the calculation of the dust mass and the characterization of the inner gaseous disk. The H$_2$CO and CS emission from the inner 200 au is mostly from the disk, and their morphology is very similar. The CN emission significantly differs from the other two molecules as it is observed only beyond 150 au. This ring-like morphology is consistent with previous observations and the predictions of thermochemical disk models. Finally, we constrained the {disk-integrated} column density of all molecules. In particular, we found that the CH$_3$OH/H$_2$CO ratio must be smaller than $\sim$2, making the methanol non-detection still consistent with the only such  ratio available from the literature (1.27 in TW Hya).}

   \keywords{astrochemistry --
                stars: pre-main sequence --
                protoplanetary disks --
                ISM: individual object: DG Tau B --
                }

\authorrunning{Garufi et al.}

\titlerunning{ALMA-DOT I. CO, H$_2$CO, CS, and CN around DG Tau B}

   \maketitle
%

\section{Introduction}
Planets form in protoplanetary disks through the assembly of dust and gas. The chemical composition of planets depends on the location and timescale for their formation and is intimately connected to the spatial distribution and abundance of the various molecular species in the disk. Over the last decade a number of chemical surveys have been carried out with either single-dish telescopes \citep[e.g.,][]{Guilloteau2012, Guilloteau2013} or low-resolution interferometers \citep[e.g.,][]{Oberg2010, Oberg2011}. The Atacama Large Millimeter Array (ALMA) has allowed the characterization of gaseous species with high spatial and spectral resolution. Much of the focus has thus far been put on the imaging of CO and its isotopologs \citep[e.g.,][]{Ansdell2016, Long2017}, although some recent surveys also helped characterize S-bearing and N-bearing molecules \citep[e.g.,][]{LeGal2019, Bergner2019}. 

On the other hand, the observations of complex organic molecules (containing more than six atoms) in protoplanetary disks are still in their early days. The characterization of these molecules is fundamental for the understanding of how genetic and metabolic molecules are formed and/or delivered to {potentially habitable planets}. To date, only simple organic molecules such as formaldehyde (H$_2$CO),  methanol (CH$_3$OH), methyl cyanide (CH$_3$CN), and formic acid (HCOOH) have been observed in protoplanetary disks and the number of such detections is very limited \citep{Oberg2015b, Favre2018}. In particular, H$_2$CO has been mapped in only   a handful of disks \citep[e.g.,][]{vanderMarel2014, Loomis2015, Oberg2017, Carney2017, Podio2019}, while the only detection of CH$_3$OH was obtained for the nearby disk of TW Hya \citep{Walsh2016}. A very important open question is whether H$_2$CO forms in the gas phase or through hydrogenation of icy molecules of CO on dust grains \citep[see, e.g.,][]{Walsh2014b}. On the other hand, methanol can only form on grains.

\begin{table*}
 \centering
 \caption{Properties of the observed lines. Columns are: molecular species, transition, frequency at rest frame ($^b$=blended line), upper-level energy, line strength, beam size, channel r.m.s., flux integrated over the main region (see text), column density. Errors on the flux have 1$\sigma$ confidence. Parentheses denote a flux above 1$\sigma$ but below 3$\sigma$ confidence. The limits on the column density are derived assuming T$\rm _{ex}$ between 30 and 300 K.}
 \label{Line_table}
  \begin{tabular}{ccccccccc}
  \hline
  Molecule & Transition & $\nu_{\rm rest}$ & E$\rm _{up}$ & S$_{ij}\mu^2$ & Beam & r.m.s. & F$_{\rm int}$ & N$_{\rm X}$ \\
   & & (GHz) & (K) & (D$^2$) & (\arcsec ) & (mJy/beam) & (mJy km/s) & (10$^{14}$ cm$^{-2}$) \\
  \hline
  \smallskip
  $^{12}$CO & 2$-$1 & 230.538000 & 17 & 0.02 & 0.14$\times$0.11 & {1.2} & - & - \\
  \smallskip
  o-H$_2$CO & $3_{1,2}-2_{1,1}$ & 225.697775 & 33 & 43.5 & 0.17$\times$0.14  & 1.4 & 622 $\pm$ 32 & 0.27 $ -$ 2.98 \\
  \smallskip
  CS & 5$-$4 & 244.935556 & 35 & 19.1 & 0.16$\times$0.13 & 0.6 & 920 $\pm$ 14 & 0.20 $-$ 0.70 \\
  \multirow{5}{*}{CN} & 2$-$1, J=$\frac{3}{2}-\frac{1}{2}$, F=$\frac{5}{2}-\frac{3}{2}$ & 226.665956 & 16 & 4.2 & \multirow{2}{*}{0.17$\times$0.14} &  \multirow{2}{*}{0.8} & \multirow{2}{*}{295 $\pm$ 18} & \multirow{2}{*}{0.17 $-$ 1.06} \\
  \smallskip
  & 2$-$1, J=$\frac{3}{2}-\frac{1}{2}$, F=$\frac{1}{2}-\frac{1}{2}$ & 226.663692$^b$ & 16 & 1.2 & & & & \\
   & 2$-$1, J=$\frac{5}{2}-\frac{3}{2}$, F=$\frac{7}{2}-\frac{5}{2}$ & 226.874781 & 16 & 6.8 & \multirow{3}{*}{0.17$\times$0.14}  &  \multirow{3}{*}{1.3}  & \multirow{3}{*}{754 $\pm$ 29}  & \multirow{3}{*}{0.18 $-$ 1.10} \\
   & 2$-$1, J=$\frac{5}{2}-\frac{3}{2}$, F=$\frac{5}{2}-\frac{3}{2}$ & 226.874190$^b$ & 16 & 4.2 & & &  &  \\
   \smallskip
   & 2$-$1, J=$\frac{5}{2}-\frac{3}{2}$, F=$\frac{3}{2}-\frac{1}{2}$ & 226.875896$^b$ & 16 & 2.6 & & & & \\
   \smallskip
   $^{34}$SO$_2$ & $4_{2,2}-3_{1,3}$ & 229.857618 & 19 & 4.6 & 0.17$\times$0.14  & 0.8 & <18 & \\
   \multirow{3}{*}{SO$_2$} & $11_{5,7}-12_{4,8}$ & 229.347630 & 122 & 3.1 & 0.17$\times$0.14  & 0.9 & (24 $\pm$ 20) & < 1.60 $-$ 4.48 \\
    & $5_{4,2}-6_{3,3}$ & 243.087647 & 53 & 0.7 & 0.16$\times$0.13  & 0.8 & (22 $\pm$ 19) & < 1.75 $-$ 11.75 \\
    \smallskip
    & $5_{2,4}-4_{1,3}$ & 241.615796 & 24 & 5.7 & 0.16$\times$0.13  & 0.9 & (42 $\pm$ 21) & < 0.16 $-$ 2.58 \\
    \multirow{2}{*}{HDO} & $3_{1,2}-2_{2,1}$ & 225.896720 & 168 & 0.7 & 0.17$\times$0.14  & 0.8 & <18 & < 0.09 $-$ 0.70 \\
    \smallskip
     & $2_{1,1}-2_{1,2}$ & 241.561550 & 95 & 0.4 & 0.16$\times$0.13  & 0.9 & <21 & < 0.24 $-$ 0.63 \\
   \multirow{2}{*}{CH$_3$OH} & $3_{-2,2}-4_{-1,4}$ & 230.027060 & 40 & 0.7 & 0.17$\times$0.14  & 0.9 & <20 & \\
     & $5_{0,5}-4_{0,4}$ & 241.791431 & 35 & 4.0 & 0.16$\times$0.13  & 0.9 & (26 $\pm$ 21) & < 0.55 $-$ 10.86 \\
   \hline
   \end{tabular}
\end{table*}

The morphological characterization of dust grains in protoplanetary disks is much more advanced than that of molecular species. In fact, high-resolution images at either near-IR or \mbox{(sub)millimeter} wavelengths are currently available for more than a hundred sources \citep[e.g.,][]{Garufi2017b, Avenhaus2018, Andrews2018, Long2018b}. The vast majority of these objects are Class II sources \citep[following][]{Lada1987}, namely stars older than $\sim$1 Myr that are visible at optical wavelengths and show a spectral energy distribution (SED) composed of a stellar blackbody plus a substantial IR excess from the surrounding disk. This census clearly reveals the systematic detection of morphological  substructures ascribed to the dynamical interaction potentially driven by embedded (unseen) planets. It is still debated whether protoplanetary disks at this stage are actually in the process of forming planets or  only host planets that have already undergone most of their accretion  \citep[see, e.g.,][]{Manara2018}. An efficient way to tackle this problem is imaging the less well studied Class I objects, namely protostars younger than 1 Myr that are deeply embedded in their natal envelope and that show a flatter SED between near- and far-IR wavelengths. The few ALMA observations of such objects \citep[e.g.,][]{ALMA2015, Sheehan2017} reveal that disk substructures are already present in the disk suggesting an early onset of planet formation. 

The observation of Class I disks at millimeter wavelengths is challenged by the contaminating emission from the residual envelope and by the prominent outflows that are often present around this type of star. Nonetheless, these early structures also carry fundamental information on the accretion and ejection processes of material between the disk and the environment. The small ALMA chemical survey of disk-outflow sources in Taurus (ALMA-DOT), based on Cycle 4 to 7 ALMA programs, is targeting {these} sources. In a pilot work, \citet{Podio2019} reported on observations of the prototypical outflow-source DG Tau showing that the high angular resolution and sensitivity of ALMA can effectively help us characterize the disk chemical composition by separating the disk emission from the outflow and envelope contamination. In this paper, we report on ALMA Cycle 4 observations of the continuum, CO, H$_2$CO, CS, and CN emission toward a prototypical Class I source \citep[e.g.,][]{Watson2004}, the low-mass star DG Tau B. 

In the visible, {the 1.1 M{$_\odot$} star} DG Tau B \citep{deValon2020} is severely obscured by its {0.07 M{$_\odot$}} circumstellar disks \citep{Guilloteau2011}, as is clear from the Hubble Space Telescope {(HST)} images by \citet{Stapelfeldt1997} and \citet{Padgett1999}. The evident dark lane visible in these images is suggestive of a strongly inclined disk. However, the millimeter observations by \citet{Guilloteau2011} constrained the disk inclination to only $\sim65\degree$, {implying that the disk opening angle for small particles in the outer regions is larger than $90\degree-65\degree=25\degree$}. In several Class II sources with more inclined disks, the star is not obscured by the disk \citep[see, e.g.,][]{Pohl2017, Langlois2018, Avenhaus2018}. This implies that the early-type SED of DG Tau B is due to its young evolutionary stage and not by a coincidental alignment of disk and line of sight. More importantly, the source is known to host a bright atomic jet \citep{Mundt1987, Podio2011} and a prominent molecular outflow \citep{Mitchell1994, Zapata2015, deValon2020}, which are indicative of a young star. CO, CN, H$_2$CO, and SO have been detected around DG Tau B by \citet{Guilloteau2013}, while CN and CO imaged with moderate angular resolution (0.5\arcsec$-$1.0\arcsec) by \citet{Guilloteau2014} and \citet{Zapata2015}, {and CO recently imaged with high angular resolution (0.15\arcsec) by \citet{deValon2020}}.

The paper is organized as follows. In Sect.\,\ref{Observations}, we describe the observing setup and the data reduction, in Sect.\,\ref{Results} we present the results of the analysis, and in Sects.\,\ref{Discussion} and \ref{Conclusions} we discuss our findings and present our conclusions. 

\section{Observations and data reduction} \label{Observations}
ALMA observations of DG Tau B were performed during \mbox{Cycle 4} in August 2017 in an extended configuration with baselines ranging from 17 m to 3.7 km. The total integration time amounted to $\sim$88 minutes. The bandpass was calibrated with the quasar J0510+1800, and phase calibration was performed every $\sim$8 minutes using the quasar J0438+3004. The correlator setup included 12 high-resolution (0.122 MHz) spectral windows (SPWs) covering 12 molecular transitions of six molecules: $^{12}$CO, H$_2$CO, CN, HDO, CH$_3$OH, and SO$_2$, plus one 1.875\,GHz broad SPW for continuum emission (and including a strong CS spectral line). 

\begin{figure*}
  \centering
 \includegraphics[width=18cm]{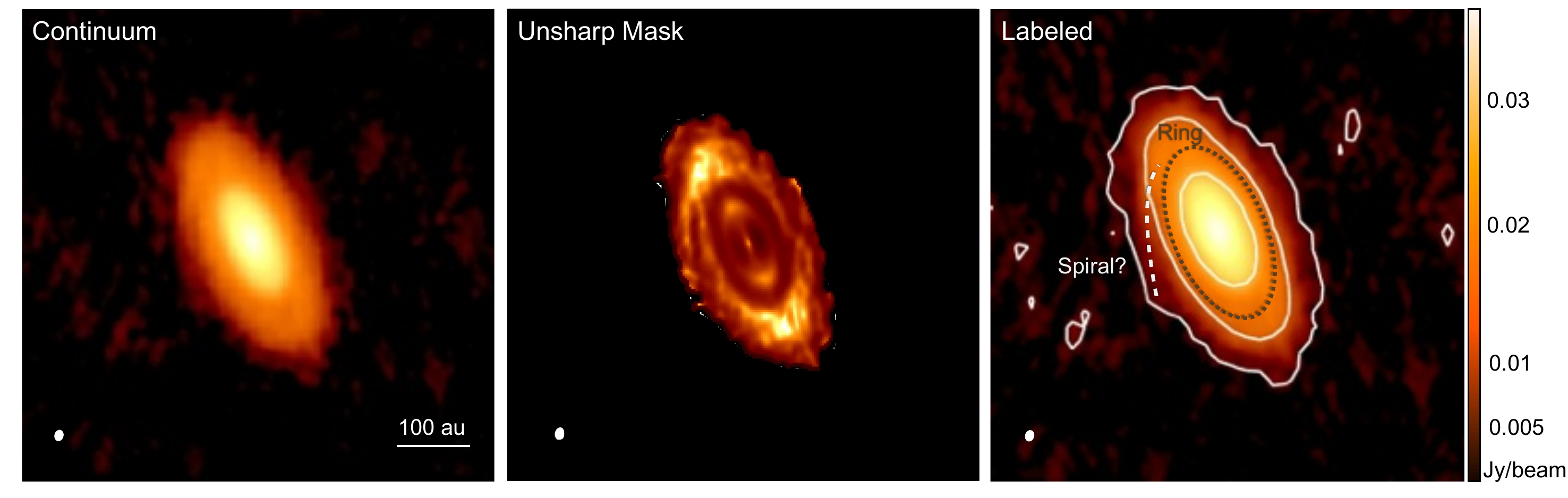} 
     \caption{Continuum emission toward DG Tau B at 235 GHz (1.27 mm). \textit{Left panel}: Original image in logarithmic color stretch. \textit{Middle panel}: Same image after applying unsharp masking, i.e.,\ the subtraction from the original image of the same image smoothed by 0.1\arcsec. \textit{Right panel}: Labeled version of the same image. The isophote curves at 3$\sigma$, 25$\sigma$, and 100$\sigma$ (with $\sigma=0.034$ mJy beam$^{-1}$) are shown together with the location of the broad ring (brown) and a putative spiral arm (gray) inferred from the unsharp masked image.} 
 \label{Continuum}
 \end{figure*}

Data reduction was carried out following standard procedures using the ALMA pipeline in CASA 4.7.2. Self-calibration was performed on the DG Tau B continuum emission, by combining a selection of line-free channels, and applying the phase-solutions to the continuum-subtracted line SPWs. {Two rounds of self-calibration were attempted with solution intervals of 180 s and 60 s. These operations improved the S/N of the continuum image from 360 to 1400 and that of the channel maps by a representative 10\%.} For a consistent astrometry among the SPWs, the same self-calibration table was applied to all the SPWs. Continuum images and spectral cubes were produced using \textsc{tclean} interactively, applying a manually selected mask, {and halting the procedure when the visual inspection of the residuals revealed no more source emission}. When imaging the spectral line cubes we used Briggs weighting of 2.0 to give more weight to the short baselines and set a channel width of 0.648 km s$^{-1}$ to improve the S/N in a single channel. {The r.m.s.\ was determined from a large area in map, in the spectral channels without detectable line emission.} The beam size and r.m.s.\ of the resulting line cubes are listed in Table \ref{Line_table}.   The maximum recoverable scale is $\sim$1.3\arcsec. The integrated intensity (moment 0) maps and intensity weighted average velocity (moment 1) maps of all lines have been obtained from channels with $V=3.4-9.4$ km s$^{-1}$, except CO from channels with $V=2.8-5.8$ km s$^{-1}$ and $V=7.0-12.4$ km s$^{-1}$ to avoid signal-free channels (see Sect.\,\ref{Gas_emission}). Moment 1 maps are obtained by clipping fluxes below 3$\sigma$.

\section{Results} \label{Results}

\subsection{Dust emission} \label{Dust_emission}
The continuum emission of DG Tau B at 235 GHz (1.27 mm) is shown in Fig.\,\ref{Continuum}. The inclined disk is clearly detected around coordinates R.A.=04:27:02.574 and Dec=26\degree 05\arcmin 30\arcsec.22. A Gaussian fit to this signal yields a position angle P.A.=23.7\degree\ east of north and inclination $i=62.3\degree$. Continuum emission is detectable above 3$\sigma$ confidence out to $\sim1.8\arcsec$. The separation encompassing 90\% of the total flux is 1.06\arcsec. This distance corresponds to the 25$\sigma$ isophote of Fig.\,\ref{Continuum} and, assuming a distance of 140 pc (see Appendix \ref{Distance_DGTauB}), translates into a physical scale of 150 au.

The continuum emission of DG Tau B reveals the presence of substructures. These are {visible} from the unsharp masked image in Fig.\,\ref{Continuum}. Unsharp masking is a technique that artificially increases the contrast of an image by means of a negative blurred version of the same image \citep[see, e.g.,][]{Garufi2016, Perez2016}. Although no quantitative analysis is recommendable from the resulting image, this procedure clearly reveals the existence of shallow rings and gaps in the dust distribution. These radial variations are also visible in the radial profile of the original image (see Sect.\,\ref{Radial_distribution}) where they can be quantified as deviations of 20-30\% in flux with respect to a constantly declining profile. In addition to the rings, a spiral-like structure is marginally visible to the  east in both the original and the unsharp masked images.

\subsection{Gas emission} \label{Gas_emission}
\begin{figure*}
  \centering
 \includegraphics[width=18.5cm]{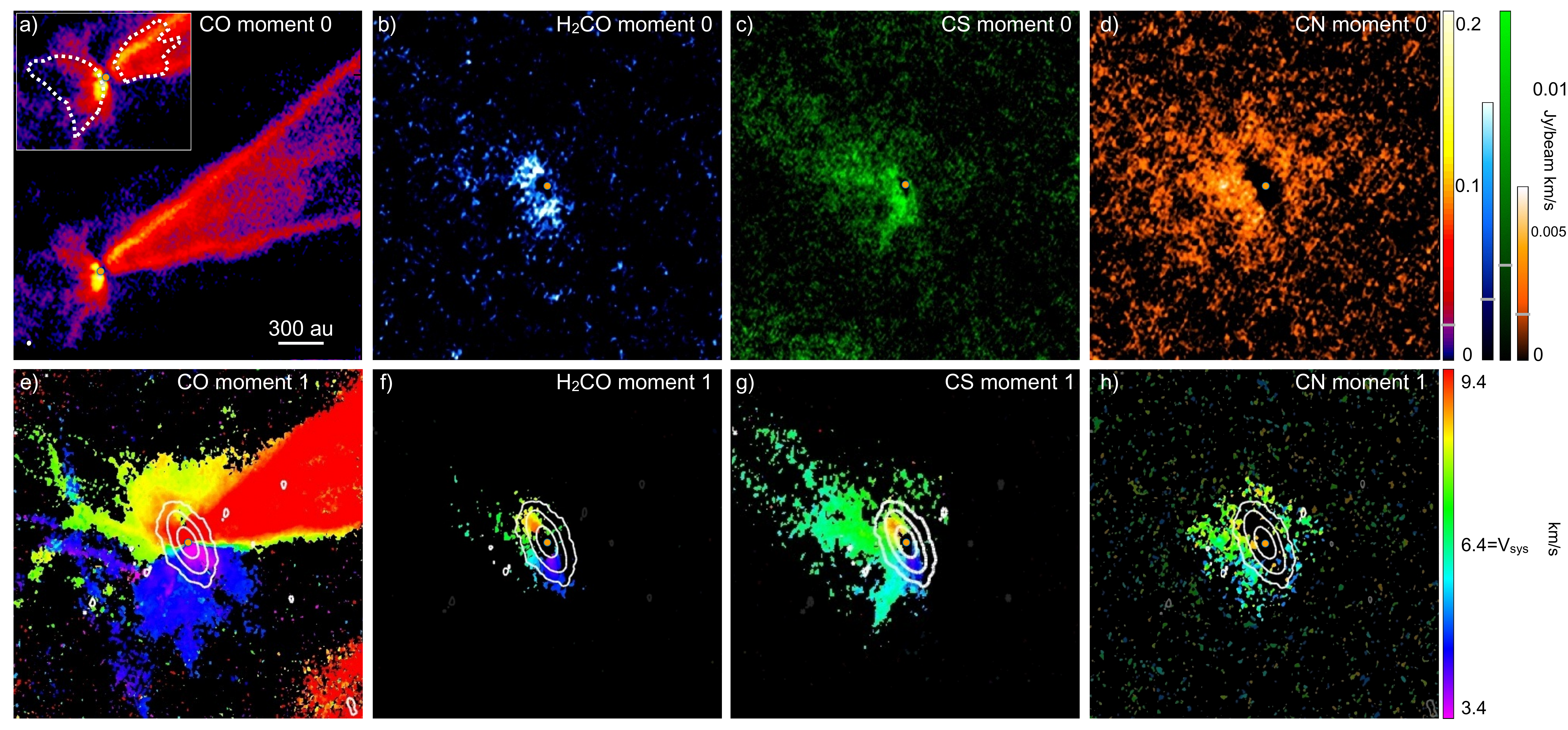} 
     \caption{Line maps from DG Tau B. \textit{(a):} CO moment 0. {Dashed contours in the inset image define the scattered light from HST \citep{Padgett1999}}. \textit{(b):} H$_2$CO moment 0.  \textit{(c):} CS moment 0. \textit{(d):} CN at 226.87 GHz moment 0. \textit{(e):} CO moment 1. \textit{(f):} H$_2$CO moment 1. \textit{(g):} CS moment 1. \textit{(h):} CN at 226.87 GHz moment 1. Fluxes at 3$\sigma$ confidence are indicated by the gray notch on the color bars. All moment 1 maps are obtained by clipping fluxes below 3$\sigma$ and are shown with the same spatial scale. Contours on these maps are the 100$\sigma$, 25$\sigma$, and 3$\sigma$ of the continuum map (Fig.\,\ref{Continuum}). The stellar position is indicated by the central orange dot (offset from the center in panel a). The beam size of all images is shown to the bottom left of (a). North is up, east is left.} 
 \label{Line_maps}
 \end{figure*}

Extended emission is revealed from the maps of four of the seven surveyed molecules:  CO, CS, CN, and H$_2$CO. Their integrated intensity maps (moment 0) and intensity-weighted velocity maps (moment 1) are shown in Fig.\,\ref{Line_maps}, while a summary of their properties can be found in Table \ref{Line_table}. It should be noted that the CN 2$-$1 transition has a hyperfine structure with 19 components ranging from 226.287 GHz to 227.191 GHz \citep[see][]{Guilloteau2013}. The spectral windows of this dataset are centered around two of the brightest hyperfine components (at 226.665 GHz and 226.874 GHz). Both lines are blended with other lines (see Table \ref{Line_table}). In this work, we focus on the latter line since its emission is brighter but similarly distributed to that of the former line.  

As is clear from Fig.\,\ref{Line_maps}, the line emission originates from two distinct physical regions:\ the outflow cavities (Sect.\,\ref{Molecular_outflow}) and the circumstellar disk (Sect.\,\ref{Circumstellar_disk}). The velocities of the detected emission span from $\sim$3 km s$^{-1}$ to $\sim$12 km s$^{-1}$. However, no CO emission is detected between 5.8 km s$^{-1}$ and 7.0 km s$^{-1}$. This velocity interval is centered around the systemic velocity of DG Tau B \citep[$\sim$6.4 km s$^{-1}$,][]{Guilloteau2013} indicating that the absence of signal is due to the absorption from the large-scale cloud around the source. The cloud would be, in turn, invisible from our data because {of spatial filtering} \citep[as in][]{Mitchell1997}. On the other hand, the CS, CN, and H$_2$CO lines are detected at the systemic velocity. 

\subsubsection{Molecular outflow} \label{Molecular_outflow}
The most prominent contribution to the CO emission derives from the two outflow cavities (Fig.\,\ref{Line_maps}a and \ref{Line_maps}e). The redshifted cone, to {the} NW, is much more extended and collimated than the blueshifted cone, to {the} SE. The former is detected from $V=7.5$ km s$^{-1}$ to 15.5 km s$^{-1}$. {The} southern and northern edges of the cone have slightly different velocities, revealing that the outflow rotates in a direction consistent with the disk rotation (see Sect.\,\ref{Circumstellar_disk}). In the inner 3\arcsec, the southern edge shows  an average velocity of 9 km s$^{-1}$, while the northern edge of nearly 10 km s$^{-1}$. The outflow width also varies with the velocity probed, spanning from 45$\degree$ at 8 km s$^{-1}$ to 30$\degree$ at 13 km s$^{-1}$. The maps of H$_2$CO, CS, and CN lines (Figs.\,\ref{Line_maps}b, \ref{Line_maps}c, and \ref{Line_maps}d) do not show any obvious counterpart to the redshifted outflow. However, a closer look to the individual frequency maps reveals marginal CS and CN flux at $V=7.5-8.5$ km s$^{-1}$ that is co-located with the CO emission (see channel maps of Fig.\,\ref{Channel_maps_2}). At higher velocities where CO becomes brighter, this CS and CN contribution is no longer visible. 

The morphology of the blueshifted outflow is far more complicated. This structure is bright in CO and, unlike the redshifted outflow, in CS (see Fig.\,\ref{Line_maps}c). Focussing on the CS only (given the above-mentioned absorption of CO at the systemic velocity), the southern and eastern walls of this outflow cavity peak at velocities of 5.8 km s$^{-1}$ and 7.0 km s$^{-1}$, respectively (see again channel maps of Fig.\,\ref{Channel_maps_1} and \ref{Channel_maps_2}). Thus, our observations resolve a rotation for the  blueshifted outflow of $\Delta V \simeq 1.2$ km s$^{-1}$, similar to the redshifted outflow. These southern and eastern outflow walls are also visible in the CN, while only the eastern wall is marginally visible in the H$_2$CO. The CS map also shows a more diffuse filament to {the} NE that is nearly parallel to the eastern wall. However, this filament peaks at slightly bluer frequencies being maximized at the systemic velocity (and being then invisible in the CO). This feature is also relatively bright in the H$_2$CO. 

Finally, our CO maps also reveal a straight, blueshifted arm to the east of the star and running along the E$-$W direction (see Fig.\,\ref{Line_maps}e and the third channel of Fig.\,\ref{Channel_maps_1}). This filament is clearly misaligned from the outflow walls and has a higher velocity with respect to the systemic velocity ($-$2 km s$^{-1}$). It could therefore represent an inflow signature. Interestingly, this structure has no counterpart in any of the other surveyed lines.

\subsubsection{Circumstellar disk} \label{Circumstellar_disk}
The gaseous circumstellar disk is bright in CO, H$_2$CO, CS, and CN emission in the inner 2\arcsec\ around DG Tau B. However, all the moment 0 maps of these lines also show an inner region with negative fluxes. This region has an aspect ratio and an orientation similar to those of the continuum emission. For H$_2$CO and CS it has a size of $\sim$0.25\arcsec\ when measured along the major axis, while it is significantly larger ($\approx 0.8\arcsec$) for CN. The same depression was imaged in CN with coarser resolution by \citet{Guilloteau2014}. The negative values of the moment 0 map are inherited from the velocities $\pm1$ km s$^{-1}$ of the systemic velocity (see line profiles in Fig.\,\ref{Line_profile}). Out of this velocity range, the flux in this region rises to zero, but never becomes positive. The only exception is CO (see Fig.\,\ref{Channel_maps_1} and \ref{Channel_maps_2}), but in this case the signal is evidently from the blueshifted outflow and not from the disk. For CN, the velocity range showing negative fluxes is larger than for CS and H$_2$CO, spanning from 4.5 to 8 km s$^{-1}$ (see Fig.\,\ref{Line_profile}). However, this feature is explained by the hyperfine nature of this transition that results in three blended lines within a small velocity shift of +0.8 and $-$1.5 km s$^{-1}$ from the main line. The origin of this inner depression is discussed in Sect.\,\ref{Depression}.

\begin{figure*}
  \centering
  \includegraphics[width=17cm]{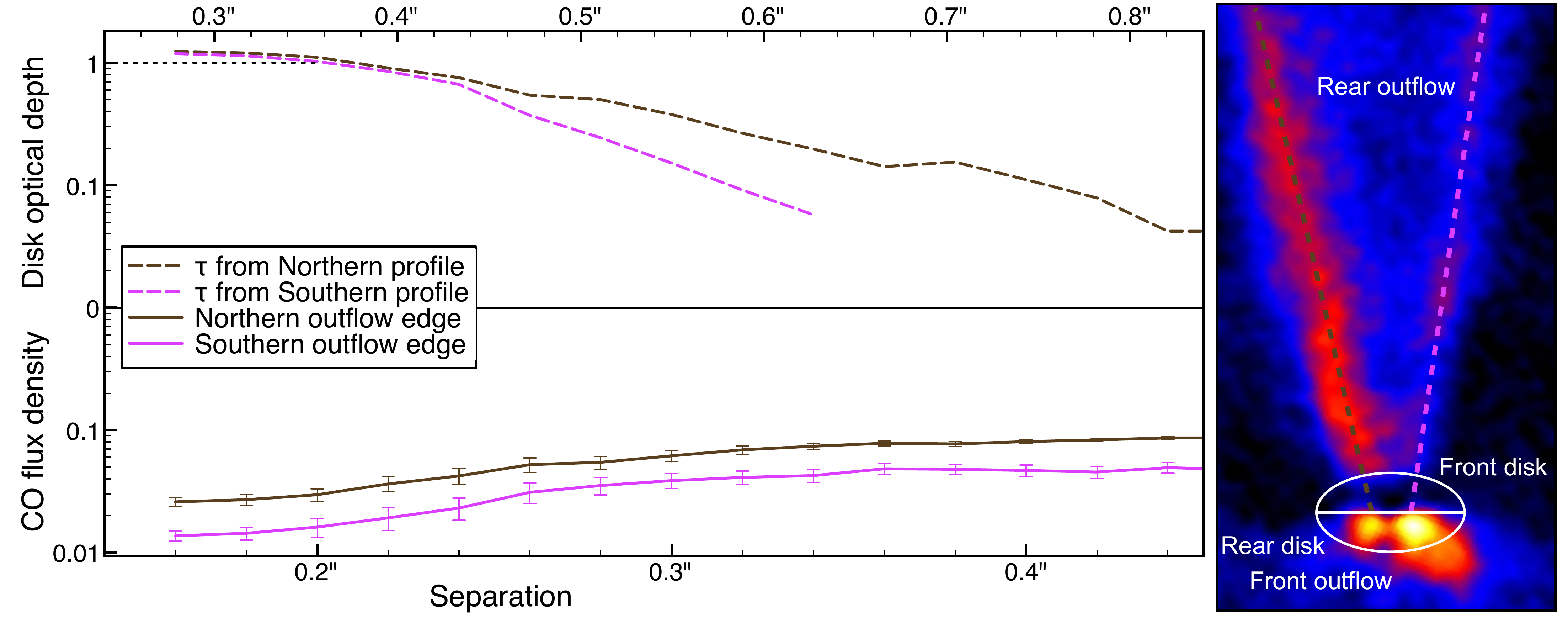} 
     \caption{Illustration of the method used to infer the inner disk optical depth. \textit{Left:} Radial profile of both outflow edges (below) and the estimated disk optical depth (above). The extent of the top panel has been corrected for the inclination. \textit{Right:} Sketch of the geometry used to extract the radial profiles.} 
 \label{Depth_image}
 \end{figure*}

From 0.25\arcsec\ to 1.0\arcsec, the CS and H$_2$CO lines show a broad spectral pattern with $\Delta V\simeq2.5$ km s$^{-1}$ around $V_{\rm sys}$; however, the former   peaks at $V_{\rm sys}$ (see Fig.\,\ref{Line_profile}) because of the strong contribution from the molecular outflow (see Sect.\,\ref{Molecular_outflow}). From Fig.\,\ref{Line_maps}f and \ref{Line_maps}g, it is clear that the distribution of both lines is shifted toward SE and is also slightly bent toward the same direction. Both effects are due to the vertical origin of this type of emission and to the disk flaring. At larger radii, the CS emission remains relatively bright and appears co-located with the CO emission. The extent of this outer disk emission can be evaluated in $\sim$3.5\arcsec\ along the disk major axis (where the outflow is absent) at $V=5.2$ and 7.6 km s$^{-1}$ (see Figs.\,\ref{Channel_maps_1} and \ref{Channel_maps_2}). On the other hand, the H$_2$CO emission becomes rapidly undetected outside 1.5\arcsec.

The distribution of the CN emission clearly differs from that of CS and H$_2$CO. First, the high-velocity components resolved for CS and H$_2$CO is undetected (see Fig.\,\ref{Line_maps}h). This is a direct consequence of the larger radii over which this molecule is not revealed. Secondly, a significant amount of flux is detected to the west of DG Tau B, where no H$_2$CO or CS flux is recovered. Most of the flux in this region has velocities between 5.2 and 7.6 km s$^{-1}$, indicating that it is emission from the disk and not from the blueshifted outflow.

\subsection{Optical depth of continuum emission} \label{Optical_depth}
The CO map reveals the presence of a faint lane at the base of the redshifted outflow. Since this outflow lies on the back of the disk, the absence of signal from the inner region is likely due to the extinction exerted by the disk. If so, we can constrain the disk optical depth from the amount of dimmed flux from the outflow. First, we extracted the radial profile of the northern and southern outflow edges, as  shown in the right panel of Fig.\,\ref{Depth_image}. Both profiles are rather flat from 0.3\arcsec\ outward, and we assumed the values of those plateaus as the unattenuated emission $I_{\rm outflow}$. Then we converted the measured intensity $I_{\rm obs}$ into the extinction optical depth $\tau_{\rm ext}$ through $\tau_{\rm ext} = \ln (I_{\rm outflow}/I_{\rm obs})$.

As is clear from the top panel of Fig.\,\ref{Depth_image}, the values of $\tau_{\rm ext}$ constrained from the two outflow edges are nearly identical inside 0.4\arcsec, while they diverge outward. This is an indication that the diminished flux of the inner region is actually due to the extinction from the disk, while at larger radii this method is biased by the intrinsic brightness variations of the two outflow edges. Thus, we conclude that the total extinction optical depth at 1.3 mm approaches unity at 0.35\arcsec. The total extinction is, in turn, given by the sum of absorption and scattering optical depths, which   may both  be important at millimeter wavelengths \citep[see][]{Isella2018}.

\subsection{Radial distribution of dust and gas emission} \label{Radial_distribution}
\begin{figure*}
  \centering
  \includegraphics[width=17cm]{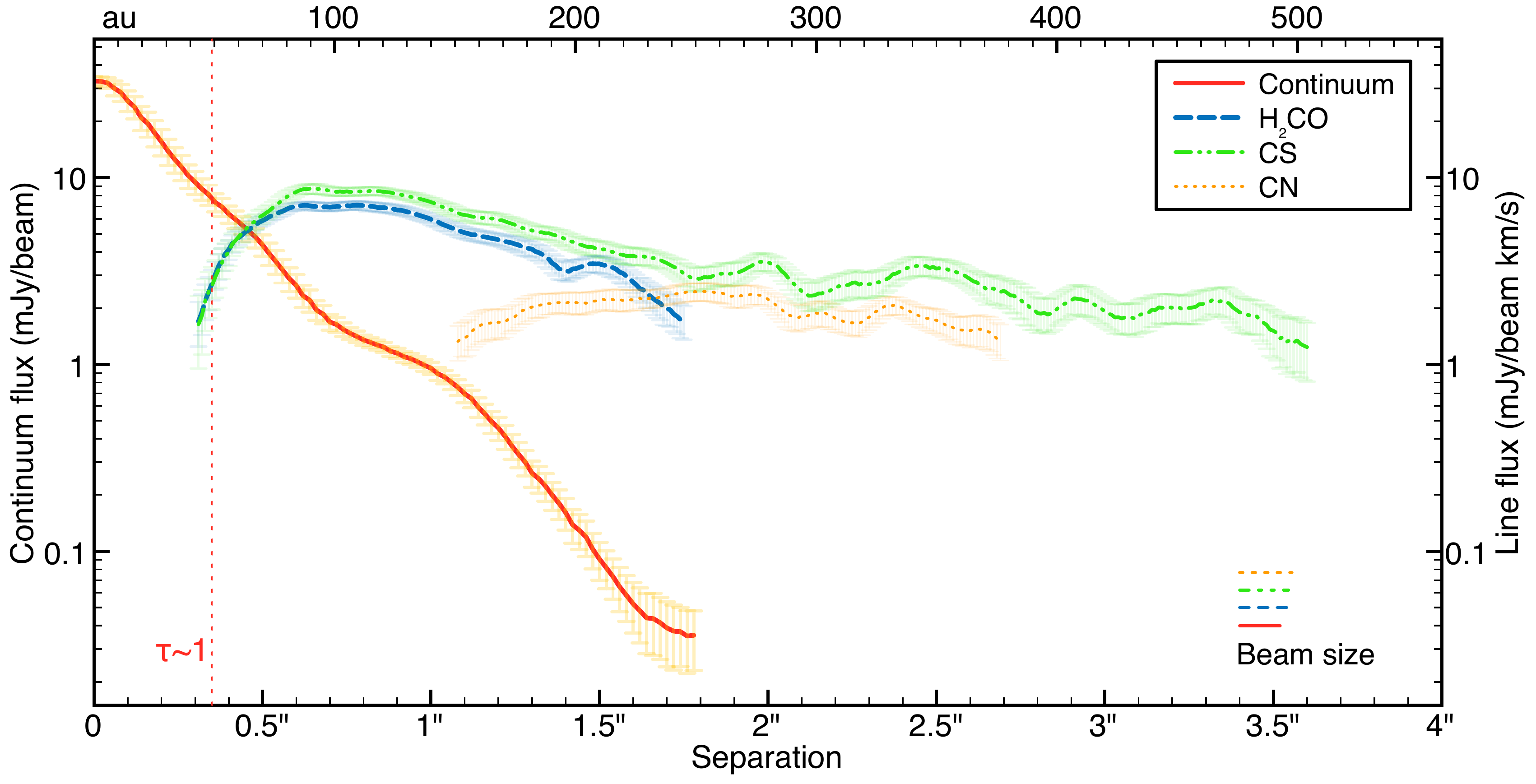} 
     \caption{Radial profile of continuum and spectral line emission, obtained from the azimuthal average of a disk with 12$\degree$ opening angle. Continuum and H$_2$CO emission is from the disk only, while that of CS and CN is also from the outflow cavity. Error bars are obtained by propagating the uncertainties at all angles and do not include the dispersion of the averaged sample. Only fluxes above 3$\sigma$ confidence are shown. The vertical line indicates the extent over which the continuum emission is optically thick (see Sect.\,\ref{Optical_depth}).} 
 \label{Radial_profile}
 \end{figure*}

A clear differentiation in the radial distribution of the dust and the various molecular species emerges from the previous sections. This trend is clear from Fig.\,\ref{Radial_profile}. The radial profiles that we show are obtained through an azimuthal average that takes into account the basic disk geometry. The choice of the average is motivated by the weakness of the molecular lines that are sometimes barely detected along specific angles (see, e.g., the CN along the major axis). Clearly, this approach does not   distinguish between the disk and outflow emission. To obtain the profile, we first extracted the continuum emission adopting the stellar position, P.A., and inclination inferred in Sect.\,\ref{Dust_emission}. To take into account the vertical origin of the molecular emission with simple geometrical considerations, we followed a technique commonly used by the near-IR community \citep{deBoer2016, Avenhaus2018} that exploits the apparent displacement from the center of any ring raised above the midplane. We superimposed the shape of projected circular rings, varying their different vertical locations until the match with the observed shift between molecular and continuum emission was reached. For both H$_2$CO and CS this exercise yielded an angle between emitting layer and midplane of 12$\degree$, corresponding to a disk scale height of 23 au at a separation of 100 au. Our poor understanding of the actual origin of CN limits the applicability of this method and, in the absence of better constraints, we also adopted 12$\degree$.

In summary, Fig.\,\ref{Radial_profile} shows the following:
\begin{itemize}
\item The dust disk is detected out to 250 au (with 90\% of the flux enclosed in 150 au, see Sect.\,\ref{Dust_emission}). The profile is much steeper than any gaseous component.
\item The dust disk shows substructures, in particular a broad ring at 140 au.
\item No line is detected at separations smaller than 50 au. This radius corresponds to the region where the dust disk is optically thick at 1.3 mm. 
\item The H$_2$CO and CS profiles are nearly identical in the inner 200 au, where the emission is mainly from the disk. At larger radii, the H$_2$CO soon becomes  undetectable, while the CS extends out to 500 au. This outer emission is partly from the disk and partly from the blueshifted outflow.
\item Unlike H$_2$CO and CS, no CN emission is detected between 50 au and 150 au. This molecule is detected out to almost 400 au, although a fraction of this emission originates, as CS, in the blueshifted outflow. 
\end{itemize}

\subsection{Fluxes of detected and undetected lines} \label{Fluxes}
To estimate the amount of large-scale flux that our interferometric observations are insensitive to, we measured the total H$_2$CO and CN 226.87 GHz flux over a region of 11\arcsec$\times$11\arcsec. In fact, \citet{Guilloteau2013} reported observations of these two lines around DG Tau B using the IRAM single-dish telescope (having a beam size equal to the selected region). We found an H$_2$CO flux of 1.25 Jy km s$^{-1}$ and CN flux of 0.93 Jy km s$^{-1}$. These values are respectively 20\% and 30\% lower than what was measured by \citet{Guilloteau2013}. In view of the 10\% uncertainty from the flux calibration, the amount of missing flux may be moderate.

A more stringent measurement of the disk flux is obtained by integrating only the flux between 0.3\arcsec\ and 2.0\arcsec\ (referred to as the main region) and considering the geometrical parameters of Sect.\,\ref{Radial_distribution}. On the other hand, the integration of the CN emission was performed between 0.8\arcsec\ and 2.5\arcsec. The resulting fluxes are shown in Table \ref{Line_table}. CS turned out to have the brightest emission, as is intuitive from Fig.\,\ref{Radial_profile}. 

None of the HDO, CH$_3$OH, and SO$_2$ lines are detected down to the achieved sensitivity of $\sim$1 mJy/beam. The total flux integrated over the main region is typically larger for SO$_2$ lines, although this flux is more than one order of magnitude lower than it is for the detected lines. We {averaged} all lines of the same species, but this approach also fails to yield any significant detection, as can be seen in the disk-integrated spectral profiles of Fig.\,\ref{Line_profile}.

\subsection{Column density of molecules} \label{Abundances}
Constraints on the  column density of a species can be obtained from the line brightness.  
Our poor knowledge of the {gas temperature and density profile does not allow us to convert the line intensity profile into a column density profile. Therefore, following the procedure described by \citet{Podio2019}, we estimated the disk-integrated column density from the line flux integrated on the line emitting area (see Sect.\,\ref{Fluxes} and fluxes in Table \ref{Line_table})}.
We assumed local thermodynamic equilibrium and optically thin emission \citep[see, e.g., Eqs.\,(1) and (2) in][]{Bianchi2017}, and adopted the molecular parameters from the Cologne Database of Molecular Spectroscopy \citep{Mueller2005}. 
{Finally, we adopted a global value of the molecular excitation temperature throughout the disk,}  leaving this parameter spanning from 30 to 300 K. The inferred column densities are summarized in Table \ref{Line_table}. 

The total column density of H$_2$CO was constrained from our o-H$_2$CO line by assuming an ortho-to-para ratio of 1.8$-$2.8 \citep{Guzman2018}, resulting in a value of $N_{\rm H_2CO}=(0.4-4.6)\cdot10^{14}$ cm$^{-2}$. From the non-detection of methanol, we then inferred a CH$_3$OH/H$_2$CO ratio < 1.5 $-$ 2.3 (depending on $T_{\rm ex}$). These limits are consistent with the only estimate available for a protoplanetary disk \citep[1.27 in TW Hya,][]{Carney2019}, and with the upper limits obtained for HD163296 and DG Tau \citep[<0.24 and <1,][]{Carney2019, Podio2019}. An approximately three times deeper exposure of the CH$_3$OH of DG Tau B is thus necessary to judge whether this ratio is comparable  to TW Hya. Among the other non-detections, the lowest limits are found for the HDO at 225.9 GHz, although these are only marginally lower than the abundances of detected species.

\section{Discussion} \label{Discussion}

\subsection{Origin of the inner depression} \label{Depression}
All detected lines show an inner depression (see Sect.\,\ref{Circumstellar_disk}). As discussed by \citet{Oberg2015}, this feature can be explained by ($i$) an actual drop in the gas surface density, ($ii$) the absorption by an optically thick disk continuum, ($iii$) the absorption by foreground material, or ($iv$) chemical reactions specific to a species that hinder its emission. 

The presence of negative values implies that the continuum absorption from foreground material is at least contributing to this inner depression. All other scenarios can at most conceal the line emission, whereas an actual absorption by colder material in the line of sight is necessary to yield negative fluxes after the process of continuum subtraction (see Sect.\,\ref{Observations}). For CS and H$_2$CO, negative fluxes are found only at the $V_{\rm sys}$, while for CN the spectral broadening is explained by the presence of blended lines (see Table \ref{Line_table}). Therefore, the absorption must be due to the rest-frame large-scale cloud. 

Within the inner depression, no H$_2$CO and CS signal is detected at any velocities. Emission from gas in Keplerian motion, however, would not be obscured by the foreground material at $V_{\rm sys}$. Thus, an additional mechanism should contribute to the inner depression. As commented in Sect.\,\ref{Radial_distribution}, the innermost CO, CS, and H$_2$CO detectable signal lies at radii closely comparable to the size of the optically thick dust continuum. This suggests that the suppression of gas emission in the inner disk region is due to the presence of optically thick, saturated continuum emission that overwhelms any line emission even if it lies closer to the observer. A similar morphology was discussed for a younger object by \citet{Harsono2018}, who invoked the substantial presence of grains larger than 1 mm to explain the high optical depth. A direct consequence of this effect is that the gaseous distribution in the inner 50 au around DG Tau B cannot be constrained from millimeter observations, but will instead require centimeter maps.

\subsection{Molecular characterization} \label{Molecular_characterization}
One of the main results of Sect.\,\ref{Results} is that the H$_2$CO and CS fluxes have similar spatial distributions and strengths in the inner 200 au (where the emission is mainly from the disk). On the other hand, the CN distribution substantially differs. In principle, this result is surprising since the emission of CS and CN is often assumed to be co-spatial \citep[see, e.g.,][]{Teague2016, Hily-Blant2017}. CN is known to be highly sensitive to UV radiation, making this molecule a good tracer of the warm disk surface \citep{Cazzoletti2018}. The greater symmetry between the near and far side with respect to CS and H$_2$CO (see Fig.\,\ref{Line_maps}) may thus support a view where (part of) the CN emission detected to the NW actually originates from the upper layer of the disk's back side. This type of detection is in fact common for relatively inclined disks, also in scattered light, since those observations trace the small dust grains at the uppermost disk layer \citep[see, e.g.,][]{Avenhaus2018}.  

\citet{Cazzoletti2018} showed that a ring-like structure \citep[as observed by, e.g.,][]{Teague2016, vanTerwisga2019} is a natural morphology of the CN emission even in full gaseous disks. Our images cannot firmly constrain such a morphology because of the inner depression discussed in Sect.\,\ref{Depression}. Nonetheless, the CN depression is much larger than that of the other molecules (see Sect.\,\ref{Radial_distribution}). This suggests that the CN is intrinsically emitted in a large ring-like structure. The models by \citet{Cazzoletti2018} showed a scaling relation between the ring width and the gaseous disk mass. Given the very high disk mass of DG Tau B, this relation may support a view where the CN emission is observed as a ring. {The range of inferred CN column densities (see Table \ref{Line_table}) is in agreement with the expectation by \citet{Cazzoletti2018}. However, the large observational uncertainty inherited from our poor knowledge of the gas temperature prevents us from  testing any scaling relations between the CN column density and the stellar and/or disk properties.} In any case, a fraction of the CN emission from DG Tau B may partly derive from the rotating envelope \citep[as concluded by][]{Guilloteau2014}.  

As for the CS, a recent small survey of TTS by \citet{LeGal2019} revealed that this emission is always centrally peaked or at most shows a small central dip. On the other hand, \citet{Teague2018b} showed an inner depression around TW Hya. There is instead a general similarity in the resolved maps of H$_2$CO, with the presence of a relatively large inner hole and strong flux being detected from the outer disk \citep{vanderMarel2014, Loomis2015, Oberg2017, Carney2017, Podio2019}. Different interpretations are given for this inner hole, with some authors leaning toward  absorption \citep[e.g.,][for HD163296]{Carney2017} and some others toward the real drop in molecular abundance \citep[e.g.,][for TW Hya and DG Tau]{Oberg2017, Podio2019}. Given the premise of Sect.\,\ref{Depression}, DG Tau B would belong to the former case. This limited sample does not allow us to find any evolutionary trend as DG Tau B and DG Tau are young sources, while HD163296 and TW Hya are older sources. None of the detected molecules show any secondary peak of emission at the edge of the millimeter dust emission, as is observed in H$_2$CO \citep{Carney2017, Oberg2017, Podio2019}, CO \citep{Huang2016}, and DCO+ \citep{Carney2018}.

H$_2$CO can either form in the gas phase from the reactions between CH$_3$ and atomic oxygen or on dust grains through double hydrogenation of the CO condensed in the grain mantle \citep{Walsh2014b, Loomis2015}. Stronger H$_2$CO fluxes from the outer disk regions are expected from the latter scenario because most of the CO is frozen onto dust grains. On the other hand, gas-phase formation is expected to produce centrally peaked emission profiles \citep{Loomis2015}. Since the inner 50 au of the gaseous component are not traced by these observations (see Sect.\,\ref{Depression}), we cannot rule out that such a centrally peaked component exists, and {thus cannot conclude whether H$_2$CO forms in gas phase or on dust grains.}

The outflow cavities of embedded objects have already been imaged in CS, CN, and H$_2$CO \citep[e.g.,][]{Bachiller2001, Codella2014, Tychoniec2019}. Intriguingly, our images reveal the detection of CS, CN, and (marginally) H$_2$CO only from the blueshifted outflow, which is however fainter in CO \citep[see also][]{deValon2020}, as well as less extended and collimated. \citet{Podio2011} inferred the same mass loss rate for the two lobes and concluded that the morphological differences must be ascribed to the different interaction for the ejecting material with an inhomogeneous environment. These diverse conditions of the medium should therefore also have an impact on the chemical conditions that allow or not the emission of CS, CN, and H$_2$CO. 

\subsection{Dust characterization} \label{Dust_characterization}
The dust disk is optically thick over approximately 20\% of its extent (even 30\% if we consider the dust outer edge of 150 au inferred in Sect.\,\ref{Dust_emission}). An important implication of this feature is that the disk dust mass estimated from millimeter observations \citep[$\approx 225\ \rm M_{\oplus}$,][]{Guilloteau2011} is appreciably underestimated. This reinforces the idea that DG Tau B is, in the context of Class I objects, an extraordinarily massive disk. In fact, its (underestimated) dust mass is four times higher than the average found for ten Class I sources in Taurus by \citet{Sheehan2017b} and three times higher than that found for three dozens Perseus sources by \citet{Tychoniec2018}. These authors also constrained the dust disk mass of Class 0 sources from the same region finding an average value comparable with that of DG Tau B. Even so, DG Tau B is only 2$-$3 times more massive than the most massive Class II disks in the young regions of Lupus (IM Lup, RU Lup), Taurus (GG Tau, GM Aur), and Chamaeleon (HD97048). This behavior is  circumstantial evidence that the dispersal of millimeter-sized grains  (by sublimation or growth) occurs less rapidly in massive disks, as also dictated by the shallow temporal trend found for these objects \citep{Garufi2018}.

Another peculiarity of the dust disk of DG Tau B is that it is  smaller than the gaseous disk \citep[see CS and CO emission at 7.0-7.6 km/s in Fig.\,\ref{Channel_maps_2} and][]{deValon2020}. In principle, this effect is suggestive of a substantial dust radial drift, although it can also be partly explained by optical depth effects \citep{Facchini2017, Facchini2019, Trapman2019}. Radial drift in DG Tau B would point to an efficient grain growth and inward migration that occurred very early in the disk lifetime given its young evolutionary stage. The efficient grain drift in massive and extended disks may be surprising since these objects are expected to generate radial traps (manifesting as rings or inner cavities) that efficiently hinder the radial drift \citep[see, e.g.,][]{Pinilla2012b}.

Finally, the dust disk of DG Tau B shows multiple rings \citep[see Fig.\,\ref{Radial_profile} and][]{deValon2020} as well as, putatively, a spiral arm (see Fig.\,\ref{Continuum}). The existence of disk substructures around a Class I object is intriguing, but it is not a first \citep[e.g.,][]{ALMA2015, Sheehan2017, Sheehan2018}. Spiral arms in young objects have already been seen in the millimeter \citep{Perez2016, Andrews2018}. This feature in DG Tau B, if confirmed, would present an interesting peculiarity. It would in fact be leading, i.e.,\ the outer region would precede the inner one. Spiral arms in another Class I object have recently been reported by \citet{Lee2019}, who concluded that the accretion from the surrounding envelope makes the outer disk gravitationally unstable and thus drives the spiral formation. {The high disk mass may potentially be sufficient to make the disk gravitationally unstable, although the disk-to-star mass ratio of DG Tau B ($\sim$0.06) is slightly lower than the formal threshold determined from analytic treatment of disk instabilities \citep[0.1; e.g.,][]{Kratter2016}.}

On the other hand, rings in disks are recurrently observed at any evolutionary stage \citep[e.g.,][]{Andrews2016, Sheehan2018}. Since the dynamical interaction with (forming) planets is among the most promising explanations for these structures \citep[e.g.,][]{Dong2015, Isella2016}, their presence so early in the disk lifetime has been used to support the scenario of rapid giant planet formation \citep{ALMA2015}. The timescale for the formation of these planets could play a role in determining the predominance of the two competing processes of dust drift and dust trapping and, in turn, could allow or hinder a rapid (1$-$2 Myr) evolution of the disk. Speculatively, in massive disks like that of DG Tau B  dust trapping eventually becomes the dominant process, and  thus  allows long disk lifetimes like those seen in TW Hya and HD163296.

\section{Conclusions} \label{Conclusions}
As part of a small ALMA chemical survey of disk-outflow sources in Taurus (ALMA-DOT), we reported on new high-resolution ($\sim$0.15\arcsec, i.e.,\ 20 au) Band 6 observations of the Class I object DG Tau B. Continuum, CO, CS, CN, and H$_2$CO emission is detected from both the circumstellar disk and the outflow cavities while SO$_2$, CH$_3$OH, and HDO lines remain undetected. The main results of the analysis are as follows:

\begin{itemize}

\item The dusty disk shows shallow substructures. In particular, a broad ring is revealed at 140 au. The image also shows marginal evidence of a spiral arm that would be leading.

\item From the occultation of the background outflow, we concluded that the inner 50 au of the disk emission is optically thick at 1.3 mm. This region corresponds to one-third of the dust disk extent (150 au) if constrained from the separation where 90\% of the flux is enclosed.

\item Emission from CO, H$_2$CO, CS, and hyperfine CN transitions is detected from our maps. In the inner 200 au the emission mostly originates from the disk while farther out it is a combination of disk and outflow signal. The distributions of the CS and H$_2$CO emission from the disk are nearly identical, while that of CN substantially differs, being relatively strong on the disk's back side. 

\item CS is detected out to 500 au and CN to 400 au, while H$_2$CO only to 250 au. As observed in other disks, the gaseous disk is therefore significantly larger than the dusty disk.

\item CO, CS, and H$_2$CO lines show a flux depression from an inner region as large as the optically thick region (50 au). The lack of line emission from this region is therefore explained by the saturated continuum emission overwhelming it. 

\item No CN signal is observed out to 150 au. This indicates that the CN is effectively emitted in a ring-like structure, in line with previous observations and modeling of this molecule. 

\item In contrast with previous observations, we do not detect a secondary line peak at the edge of the millimeter continuum.

\item The {obscuration of the inner disk region prevents us from concluding whether the H$_2$CO is formed in gas phase, which would produce centrally peaked emission, or on dust grains, which instead would enhance emission from the outer disk.}

\item We determined the upper limits to the column density of SO$_2$, CH$_3$OH, and HDO. This reveals, in particular, that the CH$_3$OH-to-H$_2$CO ratio must be lower than 2.3.

\item The CO emission from the outflow cavities is very bright. The redshifted outflow is more extended and collimated. Its aperture varies with the velocity probed.  

\item Strong CS and CN emission as well as faint H$_2$CO emission is detected from the blueshifted outflow only. The different morphology and chemistry of the two outflows suggest that the interaction of ejecting material with the medium is subject to different physical conditions.

\end{itemize}

DG Tau B hosts an exceptionally massive disk even in the context of Class I objects. The substantial dust growth and radial drift as well as the presence of shallow rings in the dust indicate a very early dust processing. These processes can respectively facilitate and be the consequence of planet formation. This could in principle suggest that planets have already formed in disks younger than 1 Myr, or at least in those as massive as the disk of DG Tau B. In any case, the large optically thick region at millimeter wavelengths inferred in this work is an important caveat to the determination of dust mass and gas column density that may also apply to other sources. 

As of today, all these results cannot be properly put into context.  DG Tau B is among the few young Class I sources that are currently well characterized and among the few protoplanetary disks with resolved maps of several chemical species. Future analogous analyses for other early-stage disks, such as  Haro 6-13 or HL Tau, will be particularly important to constrain the occurrence of these peculiarities. The ultimate question arising from this work is in fact whether all young, embedded disks will resemble (a smaller scale of) DG Tau B or whether this source can be considered  an exceptional object and  the precursor of other extraordinarily long-lived disks like those of TW Hya and HD163296.

\begin{acknowledgements}
     We are grateful to {the referee for fruitful comments} and to Alois de Valon and Catherine Dougados for useful discussions. This paper uses ALMA data from project 2016.1.00846.S. ALMA is a partnership of ESO (representing its member states), NSF (USA), and NINS (Japan), together with NRC (Canada), MOST and ASIAA (Taiwan), and KASI (Republic of Korea), in cooperation with the Republic of Chile. The Joint ALMA Observatory is operated by ESO, AUI/NRAO, and NAOJ. This work was supported by the PRIN-INAF 2016 "The Cradle of Life - GENESIS-SKA (General Conditions in Early Planetary Systems for the rise of life with SKA)", the program PRIN-MIUR 2015 STARS in the CAOS - Simulation Tools for Astrochemical Reactivity and Spectroscopy in the Cyberinfrastructure for Astrochemical Organic Species (2015F59J3R, MIUR Ministero dell'Istruzione, dell'Universit\`{a}, della Ricerca e della Scuola Normale Superiore), the European Research Council (ERC) under the European Union's Horizon 2020 research and innovation programme, for the Project "The Dawn of Organic Chemistry" (DOC), Grant No 741002, and the European MARIE SKLODOWSKA-CURIE ACTIONS under the European Union's Horizon 2020 research and innovation programme, for the Project "Astro-Chemistry Origins" (ACO), Grant No 811312. We also acknowledge support from INAF/Frontiera (Fostering high ResolutiON Technology and Innovation for Exoplanets and Research in Astrophysics) through the "Progetti Premiali" funding scheme of the Italian Ministry of Education, University, and Research, as well as NSF grants AST- 1514670 and NASA NNX16AB48G. This work was also supported by the Italian Ministero dell'Istruzione, Universit\`{a} e Ricerca through the grant Progetti Premiali 2012 - iALMA (CUP C52I13000140001), by the Deutsche Forschungs-Gemeinschaft (DFG, German Research Foundation) - Ref no. FOR 2634/1 TE 1024/1-1, by the DFG cluster of excellence ORIGINS (www.origins-cluster.de), and by the European Union's Horizon2020 research and innovation programme under the Marie Sklodowska-Curie grant agreement No 823823 (RISE DUSTBUSTERS project). D.F.\ acknowledges financial support provided by the Italian Ministry of Education, Universities and Research,project SIR(RBSI14ZRHR).
     \end{acknowledgements}

\bibliographystyle{aa} 
\bibliography{../MasterReference.bib} 

\begin{thebibliography}{73}
\expandafter\ifx\csname natexlab\endcsname\relax\def\natexlab#1{#1}\fi

\bibitem[{{ALMA Partnership} {et~al.}(2015){ALMA Partnership}, {Brogan},
  {P{\'e}rez}, {Hunter}, {Dent}, {Hales}, {Hills}, {Corder}, {Fomalont},
  {Vlahakis}, {Asaki}, {Barkats}, {Hirota}, {Hodge}, {Impellizzeri}, {Kneissl},
  {Liuzzo}, {Lucas}, {Marcelino}, {Matsushita}, {Nakanishi}, {Phillips},
  {Richards}, {Toledo}, {Aladro}, {Broguiere}, {Cortes}, {Cortes}, {Espada},
  {Galarza}, {Garcia-Appadoo}, {Guzman-Ramirez}, {Humphreys}, {Jung}, {Kameno},
  {Laing}, {Leon}, {Marconi}, {Mignano}, {Nikolic}, {Nyman}, {Radiszcz},
  {Remijan}, {Rod{\'o}n}, {Sawada}, {Takahashi}, {Tilanus}, {Vila Vilaro},
  {Watson}, {Wiklind}, {Akiyama}, {Chapillon}, {de Gregorio-Monsalvo}, {Di
  Francesco}, {Gueth}, {Kawamura}, {Lee}, {Nguyen Luong}, {Mangum}, {Pietu},
  {Sanhueza}, {Saigo}, {Takakuwa}, {Ubach}, {van Kempen}, {Wootten},
  {Castro-Carrizo}, {Francke}, {Gallardo}, {Garcia}, {Gonzalez}, {Hill},
  {Kaminski}, {Kurono}, {Liu}, {Lopez}, {Morales}, {Plarre}, {Schieven},
  {Testi}, {Videla}, {Villard}, {Andreani}, {Hibbard}, \&
  {Tatematsu}}]{ALMA2015}
{ALMA Partnership}, {Brogan}, C.~L., {P{\'e}rez}, L.~M., {et~al.} 2015, \apjl,
  808, L3

\bibitem[{{Andrews} {et~al.}(2018){Andrews}, {Huang}, {P{\'e}rez}, {Isella},
  {Dullemond}, {Kurtovic}, {Guzm{\'a}n}, {Carpenter}, {Wilner}, \&
  {Zhang}}]{Andrews2018}
{Andrews}, S.~M., {Huang}, J., {P{\'e}rez}, L.~M., {et~al.} 2018, \apj, 869,
  L41

\bibitem[{{Andrews} {et~al.}(2016){Andrews}, {Wilner}, {Zhu}, {Birnstiel},
  {Carpenter}, {P{\'e}rez}, {Bai}, {{\"O}berg}, {Hughes}, {Isella}, \&
  {Ricci}}]{Andrews2016}
{Andrews}, S.~M., {Wilner}, D.~J., {Zhu}, Z., {et~al.} 2016, \apjl, 820, L40

\bibitem[{{Ansdell} {et~al.}(2016){Ansdell}, {Williams}, {van der Marel},
  {Carpenter}, {Guidi}, {Hogerheijde}, {Mathews}, {Manara}, {Miotello},
  {Natta}, {Oliveira}, {Tazzari}, {Testi}, {van Dishoeck}, \& {van
  Terwisga}}]{Ansdell2016}
{Ansdell}, M., {Williams}, J.~P., {van der Marel}, N., {et~al.} 2016, \apj,
  828, 46

\bibitem[{{Avenhaus} {et~al.}(2018){Avenhaus}, {Quanz}, {Garufi}, {Perez},
  {Casassus}, {Pinte}, {Bertrang}, {Caceres}, {Benisty}, \&
  {Dominik}}]{Avenhaus2018}
{Avenhaus}, H., {Quanz}, S.~P., {Garufi}, A., {et~al.} 2018, The Astrophysical
  Journal, 863, 44

\bibitem[{{Bachiller} {et~al.}(2001){Bachiller}, {P{\'e}rez Guti{\'e}rrez},
  {Kumar}, \& {Tafalla}}]{Bachiller2001}
{Bachiller}, R., {P{\'e}rez Guti{\'e}rrez}, M., {Kumar}, M.~S.~N., \&
  {Tafalla}, M. 2001, \aap, 372, 899

\bibitem[{{Bergner} {et~al.}(2019){Bergner}, {{\"O}berg}, {Bergin}, {Loomis},
  {Pegues}, \& {Qi}}]{Bergner2019}
{Bergner}, J.~B., {{\"O}berg}, K.~I., {Bergin}, E.~A., {et~al.} 2019, \apj,
  876, 25

\bibitem[{{Bianchi} {et~al.}(2017){Bianchi}, {Codella}, {Ceccarelli}, {Taquet},
  {Cabrit}, {Bacciotti}, {Bachiller}, {Chapillon}, {Gueth}, {Gusdorf},
  {Lefloch}, {Leurini}, {Podio}, {Rygl}, {Tabone}, \& {Tafalla}}]{Bianchi2017}
{Bianchi}, E., {Codella}, C., {Ceccarelli}, C., {et~al.} 2017, \aap, 606, L7

\bibitem[{{Carney} {et~al.}(2018){Carney}, {Fedele}, {Hogerheijde}, {Favre},
  {Walsh}, {Bruderer}, {Miotello}, {Murillo}, {Klaassen}, {Henning}, \& {van
  Dishoeck}}]{Carney2018}
{Carney}, M.~T., {Fedele}, D., {Hogerheijde}, M.~R., {et~al.} 2018, \aap, 614,
  A106

\bibitem[{{Carney} {et~al.}(2019){Carney}, {Hogerheijde}, {Guzm{\'a}n},
  {Walsh}, {{\"O}berg}, {Fayolle}, {Cleeves}, {Carpenter}, \&
  {Qi}}]{Carney2019}
{Carney}, M.~T., {Hogerheijde}, M.~R., {Guzm{\'a}n}, V.~V., {et~al.} 2019,
  \aap, 623, A124

\bibitem[{{Carney} {et~al.}(2017){Carney}, {Hogerheijde}, {Loomis}, {Salinas},
  {{\"O}berg}, {Qi}, \& {Wilner}}]{Carney2017}
{Carney}, M.~T., {Hogerheijde}, M.~R., {Loomis}, R.~A., {et~al.} 2017, \aap,
  605, A21

\bibitem[{{Cazzoletti} {et~al.}(2018){Cazzoletti}, {van Dishoeck}, {Visser},
  {Facchini}, \& {Bruderer}}]{Cazzoletti2018}
{Cazzoletti}, P., {van Dishoeck}, E.~F., {Visser}, R., {Facchini}, S., \&
  {Bruderer}, S. 2018, \aap, 609, A93

\bibitem[{{Codella} {et~al.}(2014){Codella}, {Cabrit}, {Gueth}, {Podio},
  {Leurini}, {Bachiller}, {Gusdorf}, {Lefloch}, {Nisini}, {Tafalla}, \&
  {Yvart}}]{Codella2014}
{Codella}, C., {Cabrit}, S., {Gueth}, F., {et~al.} 2014, \aap, 568, L5

\bibitem[{{de Boer} {et~al.}(2016){de Boer}, {Salter}, {Benisty}, {Vigan},
  {Boccaletti}, {Pinilla}, {Ginski}, {Juhasz}, {Maire}, {Messina}, {Desidera},
  {Cheetham}, {Girard}, {Wahhaj}, {Langlois}, {Bonnefoy}, {Beuzit}, {Buenzli},
  {Chauvin}, {Dominik}, {Feldt}, {Gratton}, {Hagelberg}, {Isella}, {Janson},
  {Keller}, {Lagrange}, {Lannier}, {Menard}, {Mesa}, {Mouillet}, {Mugrauer},
  {Peretti}, {Perrot}, {Sissa}, {Snik}, {Vogt}, {Zurlo}, \& {SPHERE
  Consortium}}]{deBoer2016}
{de Boer}, J., {Salter}, G., {Benisty}, M., {et~al.} 2016, \aap, 595, A114

\bibitem[{{de Valon} {et~al.}(2020){de Valon}, {Dougados}, {Cabrit}, {Louvet},
  {Zapata}, \& {Mardones}}]{deValon2020}
{de Valon}, A., {Dougados}, C., {Cabrit}, S., {et~al.} 2020, arXiv e-prints,
  arXiv:2001.09776

\bibitem[{{Dong} {et~al.}(2015){Dong}, {Zhu}, \& {Whitney}}]{Dong2015}
{Dong}, R., {Zhu}, Z., \& {Whitney}, B. 2015, \apj, 809, 93

\bibitem[{{Facchini} {et~al.}(2017){Facchini}, {Birnstiel}, {Bruderer}, \& {van
  Dishoeck}}]{Facchini2017}
{Facchini}, S., {Birnstiel}, T., {Bruderer}, S., \& {van Dishoeck}, E.~F. 2017,
  \aap, 605, A16

\bibitem[{{Facchini} {et~al.}(2019){Facchini}, {van Dishoeck}, {Manara},
  {Tazzari}, {Maud}, {Cazzoletti}, {Rosotti}, {van der Marel}, {Pinilla}, \&
  {Clarke}}]{Facchini2019}
{Facchini}, S., {van Dishoeck}, E.~F., {Manara}, C.~F., {et~al.} 2019, arXiv
  e-prints, arXiv:1905.09204

\bibitem[{{Favre} {et~al.}(2018){Favre}, {Fedele}, {Semenov}, {Parfenov},
  {Codella}, {Ceccarelli}, {Bergin}, {Chapillon}, {Testi}, {Hersant},
  {Lefloch}, {Fontani}, {Blake}, {Cleeves}, {Qi}, {Schwarz}, \&
  {Taquet}}]{Favre2018}
{Favre}, C., {Fedele}, D., {Semenov}, D., {et~al.} 2018, \apjl, 862, L2

\bibitem[{{Gaia Collaboration} {et~al.}(2018){Gaia Collaboration}, {Brown},
  {Vallenari}, {Prusti}, {de Bruijne}, {Babusiaux}, {Bailer-Jones}, {Biermann},
  {Evans}, {Eyer}, {Jansen}, {Jordi}, {Klioner}, {Lammers}, {Lindegren},
  {Luri}, {Mignard}, {Panem}, {Pourbaix}, {Randich}, {Sartoretti}, {Siddiqui},
  {Soubiran}, {van Leeuwen}, {Walton}, {Arenou}, {Bastian}, {Cropper},
  {Drimmel}, {Katz}, {Lattanzi}, {Bakker}, {Cacciari}, {Casta{\~n}eda},
  {Chaoul}, {Cheek}, {De Angeli}, {Fabricius}, {Guerra}, {Holl}, {Masana},
  {Messineo}, {Mowlavi}, {Nienartowicz}, {Panuzzo}, {Portell}, {Riello},
  {Seabroke}, {Tanga}, {Th{\'e}venin}, {Gracia-Abril}, {Comoretto},
  {Garcia-Reinaldos}, {Teyssier}, {Altmann}, {Andrae}, {Audard},
  {Bellas-Velidis}, {Benson}, {Berthier}, {Blomme}, {Burgess}, {Busso},
  {Carry}, {Cellino}, {Clementini}, {Clotet}, {Creevey}, {Davidson}, {De
  Ridder}, {Delchambre}, {Dell'Oro}, {Ducourant},
  {Fern{\'a}ndez-Hern{\'a}ndez}, {Fouesneau}, {Fr{\'e}mat}, {Galluccio},
  {Garc{\'\i}a-Torres}, {Gonz{\'a}lez-N{\'u}{\~n}ez}, {Gonz{\'a}lez-Vidal},
  {Gosset}, {Guy}, {Halbwachs}, {Hambly}, {Harrison}, {Hern{\'a}ndez},
  {Hestroffer}, {Hodgkin}, {Hutton}, {Jasniewicz}, {Jean-Antoine-Piccolo},
  {Jordan}, {Korn}, {Krone-Martins}, {Lanzafame}, {Lebzelter}, {L{\"o}ffler},
  {Manteiga}, {Marrese}, {Mart{\'\i}n-Fleitas}, {Moitinho}, {Mora}, {Muinonen},
  {Osinde}, {Pancino}, {Pauwels}, {Petit}, {Recio-Blanco}, {Richards},
  {Rimoldini}, {Robin}, {Sarro}, {Siopis}, {Smith}, {Sozzetti}, {S{\"u}veges},
  {Torra}, {van Reeven}, {Abbas}, {Abreu Aramburu}, {Accart}, {Aerts},
  {Altavilla}, {{\'A}lvarez}, {Alvarez}, {Alves}, {Anderson}, {Andrei},
  {Anglada Varela}, {Antiche}, {Antoja}, {Arcay}, {Astraatmadja}, {Bach},
  {Baker}, {Balaguer-N{\'u}{\~n}ez}, {Balm}, {Barache}, {Barata}, {Barbato},
  {Barblan}, {Barklem}, {Barrado}, {Barros}, {Barstow}, {Bartholom{\'e}
  Mu{\~n}oz}, {Bassilana}, {Becciani}, {Bellazzini}, {Berihuete}, {Bertone},
  {Bianchi}, {Bienaym{\'e}}, {Blanco-Cuaresma}, {Boch}, {Boeche}, {Bombrun},
  {Borrachero}, {Bossini}, {Bouquillon}, {Bourda}, {Bragaglia}, {Bramante},
  {Breddels}, {Bressan}, {Brouillet}, {Br{\"u}semeister}, {Brugaletta},
  {Bucciarelli}, {Burlacu}, {Busonero}, {Butkevich}, {Buzzi}, {Caffau},
  {Cancelliere}, {Cannizzaro}, {Cantat-Gaudin}, {Carballo}, {Carlucci},
  {Carrasco}, {Casamiquela}, {Castellani}, {Castro-Ginard}, {Charlot},
  {Chemin}, {Chiavassa}, {Cocozza}, {Costigan}, {Cowell}, {Crifo}, {Crosta},
  {Crowley}, {Cuypers}, {Dafonte}, {Damerdji}, {Dapergolas}, {David}, {David},
  {de Laverny}, {De Luise}, {De March}, {de Martino}, {de Souza}, {de Torres},
  {Debosscher}, {del Pozo}, {Delbo}, {Delgado}, {Delgado}, {Di Matteo},
  {Diakite}, {Diener}, {Distefano}, {Dolding}, {Drazinos}, {Dur{\'a}n},
  {Edvardsson}, {Enke}, {Eriksson}, {Esquej}, {Eynard Bontemps}, {Fabre},
  {Fabrizio}, {Faigler}, {Falc{\~a}o}, {Farr{\`a}s Casas}, {Federici},
  {Fedorets}, {Fernique}, {Figueras}, {Filippi}, {Findeisen}, {Fonti},
  {Fraile}, {Fraser}, {Fr{\'e}zouls}, {Gai}, {Galleti}, {Garabato},
  {Garc{\'\i}a-Sedano}, {Garofalo}, {Garralda}, {Gavel}, {Gavras}, {Gerssen},
  {Geyer}, {Giacobbe}, {Gilmore}, {Girona}, {Giuffrida}, {Glass}, {Gomes},
  {Granvik}, {Gueguen}, {Guerrier}, {Guiraud}, {Guti{\'e}rrez-S{\'a}nchez},
  {Haigron}, {Hatzidimitriou}, {Hauser}, {Haywood}, {Heiter}, {Helmi}, {Heu},
  {Hilger}, {Hobbs}, {Hofmann}, {Holland}, {Huckle}, {Hypki}, {Icardi},
  {Jan{\ss}en}, {Jevardat de Fombelle}, {Jonker}, {Juh{\'a}sz}, {Julbe},
  {Karampelas}, {Kewley}, {Klar}, {Kochoska}, {Kohley}, {Kolenberg},
  {Kontizas}, {Kontizas}, {Koposov}, {Kordopatis}, {Kostrzewa-Rutkowska},
  {Koubsky}, {Lambert}, {Lanza}, {Lasne}, {Lavigne}, {Le Fustec}, {Le
  Poncin-Lafitte}, {Lebreton}, {Leccia}, {Leclerc}, {Lecoeur-Taibi},
  {Lenhardt}, {Leroux}, {Liao}, {Licata}, {Lindstr{\o}m}, {Lister}, {Livanou},
  {Lobel}, {L{\'o}pez}, {Managau}, {Mann}, {Mantelet}, {Marchal}, {Marchant},
  {Marconi}, {Marinoni}, {Marschalk{\'o}}, {Marshall}, {Martino}, {Marton},
  {Mary}, {Massari}, {Matijevi{\v{c}}}, {Mazeh}, {McMillan}, {Messina},
  {Michalik}, {Millar}, {Molina}, {Molinaro}, {Moln{\'a}r}, {Montegriffo},
  {Mor}, {Morbidelli}, {Morel}, {Morris}, {Mulone}, {Muraveva}, {Musella},
  {Nelemans}, {Nicastro}, {Noval}, {O'Mullane}, {Ord{\'e}novic},
  {Ord{\'o}{\~n}ez-Blanco}, {Osborne}, {Pagani}, {Pagano}, {Pailler},
  {Palacin}, {Palaversa}, {Panahi}, {Pawlak}, {Piersimoni}, {Pineau}, {Plachy},
  {Plum}, {Poggio}, {Poujoulet}, {Pr{\v{s}}a}, {Pulone}, {Racero}, {Ragaini},
  {Rambaux}, {Ramos-Lerate}, {Regibo}, {Reyl{\'e}}, {Riclet}, {Ripepi}, {Riva},
  {Rivard}, {Rixon}, {Roegiers}, {Roelens}, {Romero-G{\'o}mez}, {Rowell},
  {Royer}, {Ruiz-Dern}, {Sadowski}, {Sagrist{\`a} Sell{\'e}s}, {Sahlmann},
  {Salgado}, {Salguero}, {Sanna}, {Santana-Ros}, {Sarasso}, {Savietto},
  {Schultheis}, {Sciacca}, {Segol}, {Segovia}, {S{\'e}gransan}, {Shih},
  {Siltala}, {Silva}, {Smart}, {Smith}, {Solano}, {Solitro}, {Sordo}, {Soria
  Nieto}, {Souchay}, {Spagna}, {Spoto}, {Stampa}, {Steele},
  {Steidelm{\"u}ller}, {Stephenson}, {Stoev}, {Suess}, {Surdej}, {Szabados},
  {Szegedi-Elek}, {Tapiador}, {Taris}, {Tauran}, {Taylor}, {Teixeira},
  {Terrett}, {Teyssand ier}, {Thuillot}, {Titarenko}, {Torra Clotet}, {Turon},
  {Ulla}, {Utrilla}, {Uzzi}, {Vaillant}, {Valentini}, {Valette}, {van Elteren},
  {Van Hemelryck}, {van Leeuwen}, {Vaschetto}, {Vecchiato}, {Veljanoski},
  {Viala}, {Vicente}, {Vogt}, {von Essen}, {Voss}, {Votruba}, {Voutsinas},
  {Walmsley}, {Weiler}, {Wertz}, {Wevers}, {Wyrzykowski}, {Yoldas},
  {{\v{Z}}erjal}, {Ziaeepour}, {Zorec}, {Zschocke}, {Zucker}, {Zurbach}, \&
  {Zwitter}}]{Gaia2018}
{Gaia Collaboration}, {Brown}, A.~G.~A., {Vallenari}, A., {et~al.} 2018, \aap,
  616, A1

\bibitem[{{Garufi} {et~al.}(2018){Garufi}, {Benisty}, {Pinilla}, {Tazzari},
  {Dominik}, {Ginski}, {Henning}, {Kral}, {Langlois}, {M{\'e}nard}, {Stolker},
  {Szulagyi}, {Villenave}, \& {van der Plas}}]{Garufi2018}
{Garufi}, A., {Benisty}, M., {Pinilla}, P., {et~al.} 2018, \aap, 620, A94

\bibitem[{{Garufi} {et~al.}(2017){Garufi}, {Benisty}, {Stolker}, {Avenhaus},
  {de Boer}, {Pohl}, {Quanz}, {Dominik}, {Ginski}, {Thalmann}, {van Boekel},
  {Boccaletti}, {Henning}, {Janson}, {Salter}, {Schmid}, {Sissa}, {Langlois},
  {Beuzit}, {Chauvin}, {Mouillet}, {Augereau}, {Bazzon}, {Biller}, {Bonnefoy},
  {Buenzli}, {Cheetham}, {Daemgen}, {Desidera}, {Engler}, {Feldt}, {Girard},
  {Gratton}, {Hagelberg}, {Keller}, {Keppler}, {Kenworthy}, {Kral}, {Lopez},
  {Maire}, {Menard}, {Mesa}, {Messina}, {Meyer}, {Milli}, {Min}, {Muller},
  {Olofsson}, {Pawellek}, {Pinte}, {Szulagyi}, {Vigan}, {Wahhaj}, {Waters}, \&
  {Zurlo}}]{Garufi2017b}
{Garufi}, A., {Benisty}, M., {Stolker}, T., {et~al.} 2017, The Messenger, 169,
  32

\bibitem[{{Garufi} {et~al.}(2016){Garufi}, {Quanz}, {Schmid}, {Mulders},
  {Avenhaus}, {Boccaletti}, {Ginski}, {Langlois}, {Stolker}, {Augereau},
  {Benisty}, {Lopez}, {Dominik}, {Gratton}, {Henning}, {Janson}, {M{\'e}nard},
  {Meyer}, {Pinte}, {Sissa}, {Vigan}, {Zurlo}, {Bazzon}, {Buenzli}, {Bonnefoy},
  {Brandner}, {Chauvin}, {Cheetham}, {Cudel}, {Desidera}, {Feldt}, {Galicher},
  {Kasper}, {Lagrange}, {Lannier}, {Maire}, {Mesa}, {Mouillet}, {Peretti},
  {Perrot}, {Salter}, \& {Wildi}}]{Garufi2016}
{Garufi}, A., {Quanz}, S.~P., {Schmid}, H.~M., {et~al.} 2016, \aap, 588, A8

\bibitem[{{Guilloteau} {et~al.}(2013){Guilloteau}, {Di Folco}, {Dutrey},
  {Simon}, {Grosso}, \& {Pi{\'e}tu}}]{Guilloteau2013}
{Guilloteau}, S., {Di Folco}, E., {Dutrey}, A., {et~al.} 2013, \aap, 549, A92

\bibitem[{{Guilloteau} {et~al.}(2011){Guilloteau}, {Dutrey}, {Pi{\'e}tu}, \&
  {Boehler}}]{Guilloteau2011}
{Guilloteau}, S., {Dutrey}, A., {Pi{\'e}tu}, V., \& {Boehler}, Y. 2011, \aap,
  529, A105

\bibitem[{{Guilloteau} {et~al.}(2012){Guilloteau}, {Dutrey}, {Wakelam},
  {Hersant}, {Semenov}, {Chapillon}, {Henning}, \&
  {Pi{\'e}tu}}]{Guilloteau2012}
{Guilloteau}, S., {Dutrey}, A., {Wakelam}, V., {et~al.} 2012, \aap, 548, A70

\bibitem[{{Guilloteau} {et~al.}(2014){Guilloteau}, {Simon}, {Pi{\'e}tu}, {Di
  Folco}, {Dutrey}, {Prato}, \& {Chapillon}}]{Guilloteau2014}
{Guilloteau}, S., {Simon}, M., {Pi{\'e}tu}, V., {et~al.} 2014, \aap, 567, A117

\bibitem[{{Guzm{\'a}n} {et~al.}(2018){Guzm{\'a}n}, {{\"O}berg}, {Carpenter},
  {Le Gal}, {Qi}, \& {Pagues}}]{Guzman2018}
{Guzm{\'a}n}, V.~V., {{\"O}berg}, K.~I., {Carpenter}, J., {et~al.} 2018, \apj,
  864, 170

\bibitem[{{Harsono} {et~al.}(2018){Harsono}, {Bjerkeli}, {van der Wiel},
  {Ramsey}, {Maud}, {Kristensen}, \& {J{\o}rgensen}}]{Harsono2018}
{Harsono}, D., {Bjerkeli}, P., {van der Wiel}, M. H.~D., {et~al.} 2018, Nature
  Astronomy, 2, 646

\bibitem[{{Hily-Blant} {et~al.}(2017){Hily-Blant}, {Magalhaes}, {Kastner},
  {Faure}, {Forveille}, \& {Qi}}]{Hily-Blant2017}
{Hily-Blant}, P., {Magalhaes}, V., {Kastner}, J., {et~al.} 2017, \aap, 603, L6

\bibitem[{{Huang} {et~al.}(2016){Huang}, {{\"O}berg}, \& {Andrews}}]{Huang2016}
{Huang}, J., {{\"O}berg}, K.~I., \& {Andrews}, S.~M. 2016, \apjl, 823, L18

\bibitem[{Isella {et~al.}(2016)Isella, Guidi, Testi, Liu, Li, Li, Weaver,
  Boehler, Carperter, De~Gregorio-Monsalvo, Manara, Natta, P\'erez, Ricci,
  Sargent, Tazzari, \& Turner}]{Isella2016}
Isella, A., Guidi, G., Testi, L., {et~al.} 2016, Phys. Rev. Lett., 117, 251101

\bibitem[{{Isella} {et~al.}(2018){Isella}, {Huang}, {Andrews}, {Dullemond},
  {Birnstiel}, {Zhang}, {Zhu}, {Guzm{\'a}n}, {P{\'e}rez}, {Bai}, {Benisty},
  {Carpenter}, {Ricci}, \& {Wilner}}]{Isella2018}
{Isella}, A., {Huang}, J., {Andrews}, S.~M., {et~al.} 2018, \apjl, 869, L49

\bibitem[{{Kratter} \& {Lodato}(2016)}]{Kratter2016}
{Kratter}, K. \& {Lodato}, G. 2016, \araa, 54, 271

\bibitem[{{Lada}(1987)}]{Lada1987}
{Lada}, C.~J. 1987, in IAU Symposium, Vol. 115, Star Forming Regions, ed.
  M.~{Peimbert} \& J.~{Jugaku}, 1--17

\bibitem[{{Langlois} {et~al.}(2018){Langlois}, {Pohl}, {Lagrange}, {Maire},
  {Mesa}, {Boccaletti}, {Gratton}, {Denneulin}, {Klahr}, {Vigan}, {Benisty},
  {Dominik}, {Bonnefoy}, {Menard}, {Avenhaus}, {Cheetham}, {Van Boekel}, {de
  Boer}, {Chauvin}, {Desidera}, {Feldt}, {Galicher}, {Ginski}, {Girard},
  {Henning}, {Janson}, {Kopytova}, {Kral}, {Ligi}, {Messina}, {Peretti},
  {Pinte}, {Sissa}, {Stolker}, {Zurlo}, {Magnard}, {Blanchard}, {Buey},
  {Suarez}, {Cascone}, {Moller-Nilsson}, {Weber}, {Petit}, \&
  {Pragt}}]{Langlois2018}
{Langlois}, M., {Pohl}, A., {Lagrange}, A.~M., {et~al.} 2018, \aap, 614, A88

\bibitem[{{Le Gal} {et~al.}(2019){Le Gal}, {{\"O}berg}, {Loomis}, {Pegues}, \&
  {Bergner}}]{LeGal2019}
{Le Gal}, R., {{\"O}berg}, K.~I., {Loomis}, R.~A., {Pegues}, J., \& {Bergner},
  J.~B. 2019, \apj, 876, 72

\bibitem[{{Lee} {et~al.}(2019){Lee}, {Li}, \& {Turner}}]{Lee2019}
{Lee}, C.-F., {Li}, Z.-Y., \& {Turner}, N.~J. 2019, Nature Astronomy, 466

\bibitem[{{Long} {et~al.}(2017){Long}, {Herczeg}, {Pascucci}, {Drabek-Maunder},
  {Mohanty}, {Testi}, {Apai}, {Hendler}, {Henning}, {Manara}, \&
  {Mulders}}]{Long2017}
{Long}, F., {Herczeg}, G.~J., {Pascucci}, I., {et~al.} 2017, \apj, 844, 99

\bibitem[{{Long} {et~al.}(2018){Long}, {Pinilla}, {Herczeg}, {Harsono},
  {Dipierro}, {Pascucci}, {Hendler}, {Tazzari}, {Ragusa}, {Salyk}, {Edwards},
  {Lodato}, {van de Plas}, {Johnstone}, {Liu}, {Boehler}, {Cabrit}, {Manara},
  {Menard}, {Mulders}, {Nisini}, {Fischer}, {Rigliaco}, {Banzatti}, {Avenhaus},
  \& {Gully-Santiago}}]{Long2018b}
{Long}, F., {Pinilla}, P., {Herczeg}, G.~J., {et~al.} 2018, \apj, 869, 17

\bibitem[{{Loomis} {et~al.}(2015){Loomis}, {Cleeves}, {{\"O}berg}, {Guzman}, \&
  {Andrews}}]{Loomis2015}
{Loomis}, R.~A., {Cleeves}, L.~I., {{\"O}berg}, K.~I., {Guzman}, V.~V., \&
  {Andrews}, S.~M. 2015, \apj, 809, L25

\bibitem[{{Manara} {et~al.}(2018){Manara}, {Morbidelli}, \&
  {Guillot}}]{Manara2018}
{Manara}, C.~F., {Morbidelli}, A., \& {Guillot}, T. 2018, \aap, 618, L3

\bibitem[{{Mitchell} {et~al.}(1994){Mitchell}, {Hasegawa}, {Dent}, \&
  {Matthews}}]{Mitchell1994}
{Mitchell}, G.~F., {Hasegawa}, T.~I., {Dent}, W. R.~F., \& {Matthews}, H.~E.
  1994, \apj, 436, L177

\bibitem[{{Mitchell} {et~al.}(1997){Mitchell}, {Sargent}, \&
  {Mannings}}]{Mitchell1997}
{Mitchell}, G.~F., {Sargent}, A.~I., \& {Mannings}, V. 1997, \apj, 483, L127

\bibitem[{{M{\"u}ller} {et~al.}(2005){M{\"u}ller}, {Schl{\"o}der}, {Stutzki},
  \& {Winnewisser}}]{Mueller2005}
{M{\"u}ller}, H. S.~P., {Schl{\"o}der}, F., {Stutzki}, J., \& {Winnewisser}, G.
  2005, Journal of Molecular Structure, 742, 215

\bibitem[{{Mundt} {et~al.}(1987){Mundt}, {Brugel}, \& {Buehrke}}]{Mundt1987}
{Mundt}, R., {Brugel}, E.~W., \& {Buehrke}, T. 1987, \apj, 319, 275

\bibitem[{{{\"O}berg} {et~al.}(2015{\natexlab{a}}){{\"O}berg}, {Furuya},
  {Loomis}, {Aikawa}, {Andrews}, {Qi}, {van Dishoeck}, \& {Wilner}}]{Oberg2015}
{{\"O}berg}, K.~I., {Furuya}, K., {Loomis}, R., {et~al.} 2015{\natexlab{a}},
  \apj, 810, 112

\bibitem[{{{\"O}berg} {et~al.}(2015{\natexlab{b}}){{\"O}berg}, {Guzm{\'a}n},
  {Furuya}, {Qi}, {Aikawa}, {Andrews}, {Loomis}, \& {Wilner}}]{Oberg2015b}
{{\"O}berg}, K.~I., {Guzm{\'a}n}, V.~V., {Furuya}, K., {et~al.}
  2015{\natexlab{b}}, \nat, 520, 198

\bibitem[{{{\"O}berg} {et~al.}(2017){{\"O}berg}, {Guzm{\'a}n}, {Merchantz},
  {Qi}, {Andrews}, {Cleeves}, {Huang}, {Loomis}, {Wilner}, \&
  {Brinch}}]{Oberg2017}
{{\"O}berg}, K.~I., {Guzm{\'a}n}, V.~V., {Merchantz}, C.~J., {et~al.} 2017,
  \apj, 839, 43

\bibitem[{{{\"O}berg} {et~al.}(2010){{\"O}berg}, {Qi}, {Fogel}, {Bergin},
  {Andrews}, {Espaillat}, {van Kempen}, {Wilner}, \& {Pascucci}}]{Oberg2010}
{{\"O}berg}, K.~I., {Qi}, C., {Fogel}, J. K.~J., {et~al.} 2010, \apj, 720, 480

\bibitem[{{{\"O}berg} {et~al.}(2011){{\"O}berg}, {Qi}, {Fogel}, {Bergin},
  {Andrews}, {Espaillat}, {Wilner}, {Pascucci}, \& {Kastner}}]{Oberg2011}
{{\"O}berg}, K.~I., {Qi}, C., {Fogel}, J. K.~J., {et~al.} 2011, \apj, 734, 98

\bibitem[{{Padgett} {et~al.}(1999){Padgett}, {Brandner}, {Stapelfeldt},
  {Strom}, {Terebey}, \& {Koerner}}]{Padgett1999}
{Padgett}, D.~L., {Brandner}, W., {Stapelfeldt}, K.~R., {et~al.} 1999, \aj,
  117, 1490

\bibitem[{{P{\'e}rez} {et~al.}(2016){P{\'e}rez}, {Carpenter}, {Andrews},
  {Ricci}, {Isella}, {Linz}, {Sargent}, {Wilner}, {Henning}, {Deller},
  {Chandler}, {Dullemond}, {Lazio}, {Menten}, {Corder}, {Storm}, {Testi},
  {Tazzari}, {Kwon}, {Calvet}, {Greaves}, {Harris}, \& {Mundy}}]{Perez2016}
{P{\'e}rez}, L.~M., {Carpenter}, J.~M., {Andrews}, S.~M., {et~al.} 2016,
  Science, 353, 1519

\bibitem[{{Pinilla} {et~al.}(2012){Pinilla}, {Birnstiel}, {Ricci}, {Dullemond},
  {Uribe}, {Testi}, \& {Natta}}]{Pinilla2012b}
{Pinilla}, P., {Birnstiel}, T., {Ricci}, L., {et~al.} 2012, \aap, 538, A114

\bibitem[{{Podio} {et~al.}(2019){Podio}, {Bacciotti}, {Fedele}, {Favre},
  {Codella}, {Rygl}, {Kamp}, {Guidi}, {Bianchi}, \& {Ceccarelli}}]{Podio2019}
{Podio}, L., {Bacciotti}, F., {Fedele}, D., {et~al.} 2019, \aap, 623, L6

\bibitem[{{Podio} {et~al.}(2011){Podio}, {Eisl{\"o}ffel}, {Melnikov}, {Hodapp},
  \& {Bacciotti}}]{Podio2011}
{Podio}, L., {Eisl{\"o}ffel}, J., {Melnikov}, S., {Hodapp}, K.~W., \&
  {Bacciotti}, F. 2011, \aap, 527, A13

\bibitem[{{Pohl} {et~al.}(2017){Pohl}, {Benisty}, {Pinilla}, {Ginski}, {de
  Boer}, {Avenhaus}, {Henning}, {Zurlo}, {Boccaletti}, {Augereau}, {Birnstiel},
  {Dominik}, {Facchini}, {Fedele}, {Janson}, {Keppler}, {Kral}, {Langlois},
  {Ligi}, {Maire}, {M{\'e}nard}, {Meyer}, {Pinte}, {Quanz}, {Sauvage},
  {Sezestre}, {Stolker}, {Szul{\'a}gyi}, {van Boekel}, {van der Plas},
  {Villenave}, {Baruffolo}, {Baudoz}, {Le Mignant}, {Maurel}, {Ramos}, \&
  {Weber}}]{Pohl2017}
{Pohl}, A., {Benisty}, M., {Pinilla}, P., {et~al.} 2017, \apj, 850, 52

\bibitem[{{Rodr{\'\i}guez} {et~al.}(2012){Rodr{\'\i}guez}, {Dzib}, {Loinard},
  {Zapata}, {Raga}, {Cant{\'o}}, \& {Riera}}]{Rodriguez2012}
{Rodr{\'\i}guez}, L.~F., {Dzib}, S.~A., {Loinard}, L., {et~al.} 2012, \rmxaa,
  48, 243

\bibitem[{{Sheehan} \& {Eisner}(2017{\natexlab{a}})}]{Sheehan2017b}
{Sheehan}, P.~D. \& {Eisner}, J.~A. 2017{\natexlab{a}}, \apj, 851, 45

\bibitem[{{Sheehan} \& {Eisner}(2017{\natexlab{b}})}]{Sheehan2017}
{Sheehan}, P.~D. \& {Eisner}, J.~A. 2017{\natexlab{b}}, \apjl, 840, L12

\bibitem[{{Sheehan} \& {Eisner}(2018)}]{Sheehan2018}
{Sheehan}, P.~D. \& {Eisner}, J.~A. 2018, \apj, 857, 18

\bibitem[{{Stapelfeldt} {et~al.}(1997){Stapelfeldt}, {Burrows}, {Krist}, \&
  {WFPC2 Science Team}}]{Stapelfeldt1997}
{Stapelfeldt}, K., {Burrows}, C.~J., {Krist}, J.~E., \& {WFPC2 Science Team}.
  1997, in IAU Symposium, Vol. 182, Herbig-Haro Flows and the Birth of Stars,
  ed. B.~{Reipurth} \& C.~{Bertout}, 355--364

\bibitem[{{Teague} {et~al.}(2016){Teague}, {Guilloteau}, {Semenov}, {Henning},
  {Dutrey}, {Pi{\'e}tu}, {Birnstiel}, {Chapillon}, {Hollenbach}, \&
  {Gorti}}]{Teague2016}
{Teague}, R., {Guilloteau}, S., {Semenov}, D., {et~al.} 2016, \aap, 592, A49

\bibitem[{{Teague} {et~al.}(2018){Teague}, {Henning}, {Guilloteau}, {Bergin},
  {Semenov}, {Dutrey}, {Flock}, {Gorti}, \& {Birnstiel}}]{Teague2018b}
{Teague}, R., {Henning}, T., {Guilloteau}, S., {et~al.} 2018, \apj, 864, 133

\bibitem[{{Trapman} {et~al.}(2019){Trapman}, {Facchini}, {Hogerheijde}, {van
  Dishoeck}, \& {Bruderer}}]{Trapman2019}
{Trapman}, L., {Facchini}, S., {Hogerheijde}, M.~R., {van Dishoeck}, E.~F., \&
  {Bruderer}, S. 2019, arXiv e-prints, arXiv:1903.06190

\bibitem[{{Tychoniec} {et~al.}(2019){Tychoniec}, {Hull}, {Kristensen}, {Tobin},
  {Le Gouellec}, \& {van Dishoeck}}]{Tychoniec2019}
{Tychoniec}, {\L}., {Hull}, C. L.~H., {Kristensen}, L.~E., {et~al.} 2019, arXiv
  e-prints, arXiv:1910.07857

\bibitem[{{Tychoniec} {et~al.}(2018){Tychoniec}, {Tobin}, {Karska}, {Chand
  ler}, {Dunham}, {Harris}, {Kratter}, {Li}, {Looney}, {Melis}, {P{\'e}rez},
  {Sadavoy}, {Segura-Cox}, \& {van Dishoeck}}]{Tychoniec2018}
{Tychoniec}, {\L}., {Tobin}, J.~J., {Karska}, A., {et~al.} 2018, \apjs, 238, 19

\bibitem[{{van der Marel} {et~al.}(2014){van der Marel}, {van Dishoeck},
  {Bruderer}, \& {van Kempen}}]{vanderMarel2014}
{van der Marel}, N., {van Dishoeck}, E.~F., {Bruderer}, S., \& {van Kempen},
  T.~A. 2014, \aap, 563, A113

\bibitem[{{van Terwisga} {et~al.}(2019){van Terwisga}, {van Dishoeck},
  {Cazzoletti}, {Facchini}, {Trapman}, {Williams}, {Manara}, {Miotello}, {van
  der Marel}, \& {Ansdell}}]{vanTerwisga2019}
{van Terwisga}, S.~E., {van Dishoeck}, E.~F., {Cazzoletti}, P., {et~al.} 2019,
  \aap, 623, A150

\bibitem[{{Walsh} {et~al.}(2016){Walsh}, {Loomis}, {{\"O}berg}, {Kama}, {van 't
  Hoff}, {Millar}, {Aikawa}, {Herbst}, {Widicus Weaver}, \&
  {Nomura}}]{Walsh2016}
{Walsh}, C., {Loomis}, R.~A., {{\"O}berg}, K.~I., {et~al.} 2016, \apjl, 823,
  L10

\bibitem[{{Walsh} {et~al.}(2014){Walsh}, {Millar}, {Nomura}, {Herbst}, {Widicus
  Weaver}, {Aikawa}, {Laas}, \& {Vasyunin}}]{Walsh2014b}
{Walsh}, C., {Millar}, T.~J., {Nomura}, H., {et~al.} 2014, \aap, 563, A33

\bibitem[{{Watson} {et~al.}(2004){Watson}, {Kemper}, {Calvet}, {Keller},
  {Furlan}, {Hartmann}, {Forrest}, {Chen}, {Uchida}, {Green}, {Sargent},
  {Sloan}, {Herter}, {Brandl}, {Houck}, {Najita}, {D'Alessio}, {Myers},
  {Barry}, {Hall}, \& {Morris}}]{Watson2004}
{Watson}, D.~M., {Kemper}, F., {Calvet}, N., {et~al.} 2004, \apjs, 154, 391

\bibitem[{{Zapata} {et~al.}(2015){Zapata}, {Lizano}, {Rodr{\'\i}guez}, {Ho},
  {Loinard}, {Fern{\'a}ndez-L{\'o}pez}, \& {Tafoya}}]{Zapata2015}
{Zapata}, L.~A., {Lizano}, S., {Rodr{\'\i}guez}, L.~F., {et~al.} 2015, \apj,
  798, 131

\end{thebibliography}

\begin{appendix}

\section{Distance to DG Tau B} \label{Distance_DGTauB}
The distance to DG Tau B is unconstrained because its embedded nature impedes the measurement of its parallax in the visible. A natural approach would be to assume the same distance as the neighboring DG Tau \citep[$d=121$ pc,][]{Gaia2018}. Their systemic velocities are comparable \citep[see][]{Guilloteau2013}, but several other Taurus sources show similar values.

\citet{Rodriguez2012} determined the proper motion of DG Tau B from VLA observations at radio wavelengths ($\Delta$R.A.\,= 3.8 $\pm$ 1.9 mas, $\Delta$Dec = $-$20.6 $\pm$ 3.3 mas). This value is comparable to but not consistent with that of DG Tau ($\Delta$R.A.\,= 6.2 $\pm$ 0.4 mas, $\Delta$Dec = $-$19.3 $\pm$ 0.2 mas). Thus, we searched for other apparently nearby stars ($<$10 pc at a reasonable 140 pc) with proper motion consistent with DG Tau B, and found in particular three: V* FV Tau/c, KPNO-Tau 13, and 2MASS J0426+2443. These objects have a \textit{Gaia} distance of 140, 133, and 119 pc. In particular, the first target shows nearly the same proper motion of DG Tau B and is  less than 1 pc away from DG Tau b. In this work, without better constraints, we assume the distance to DG Tau B to be 140 pc. However, we cannot rule out that the source is closer to DG Tau and 2MASS J0426+2443 at \mbox{$d=$120 pc}. 

\section{Channel maps and line profiles}
\begin{figure*}
  \centering
 \includegraphics[width=6cm]{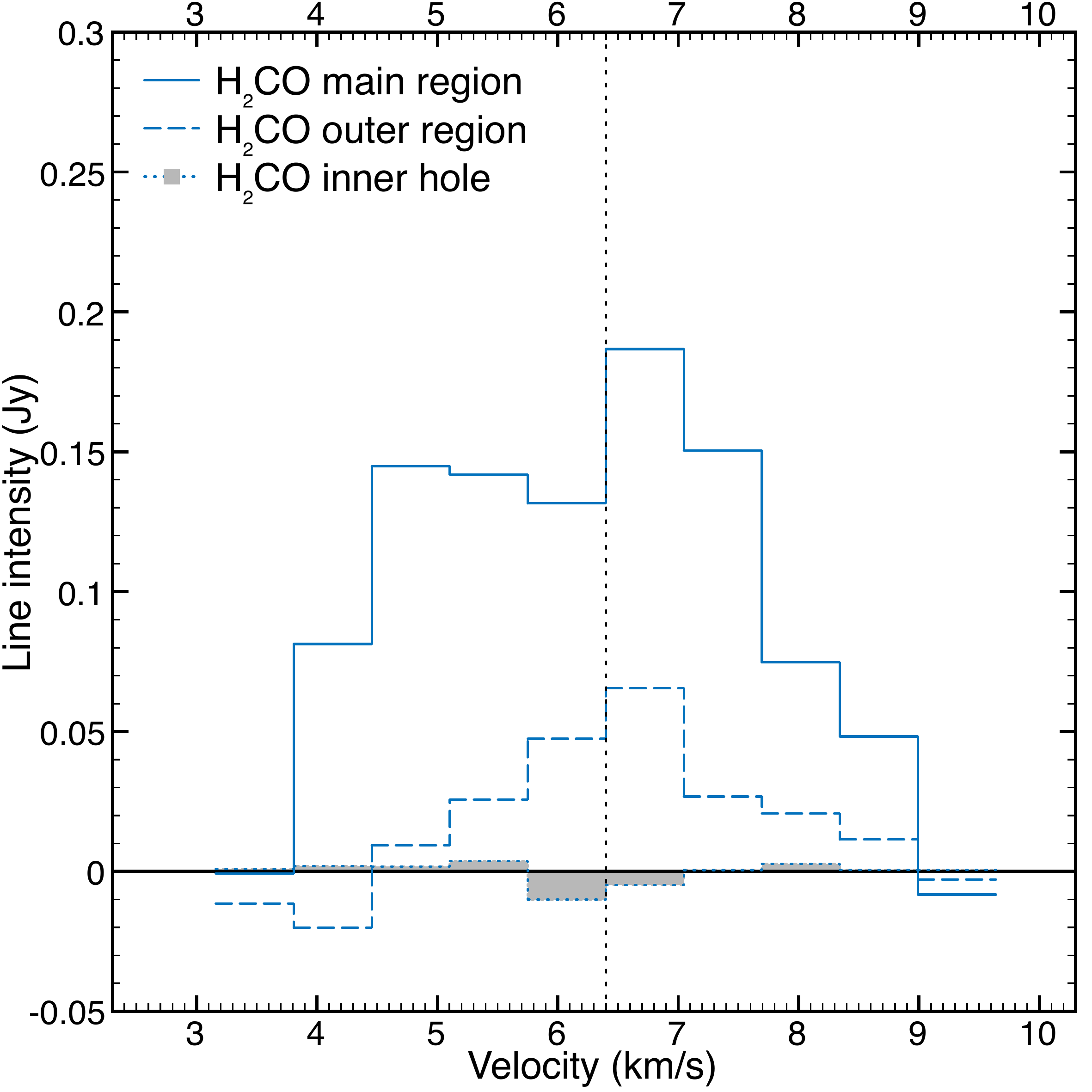}
  \includegraphics[width=6cm]{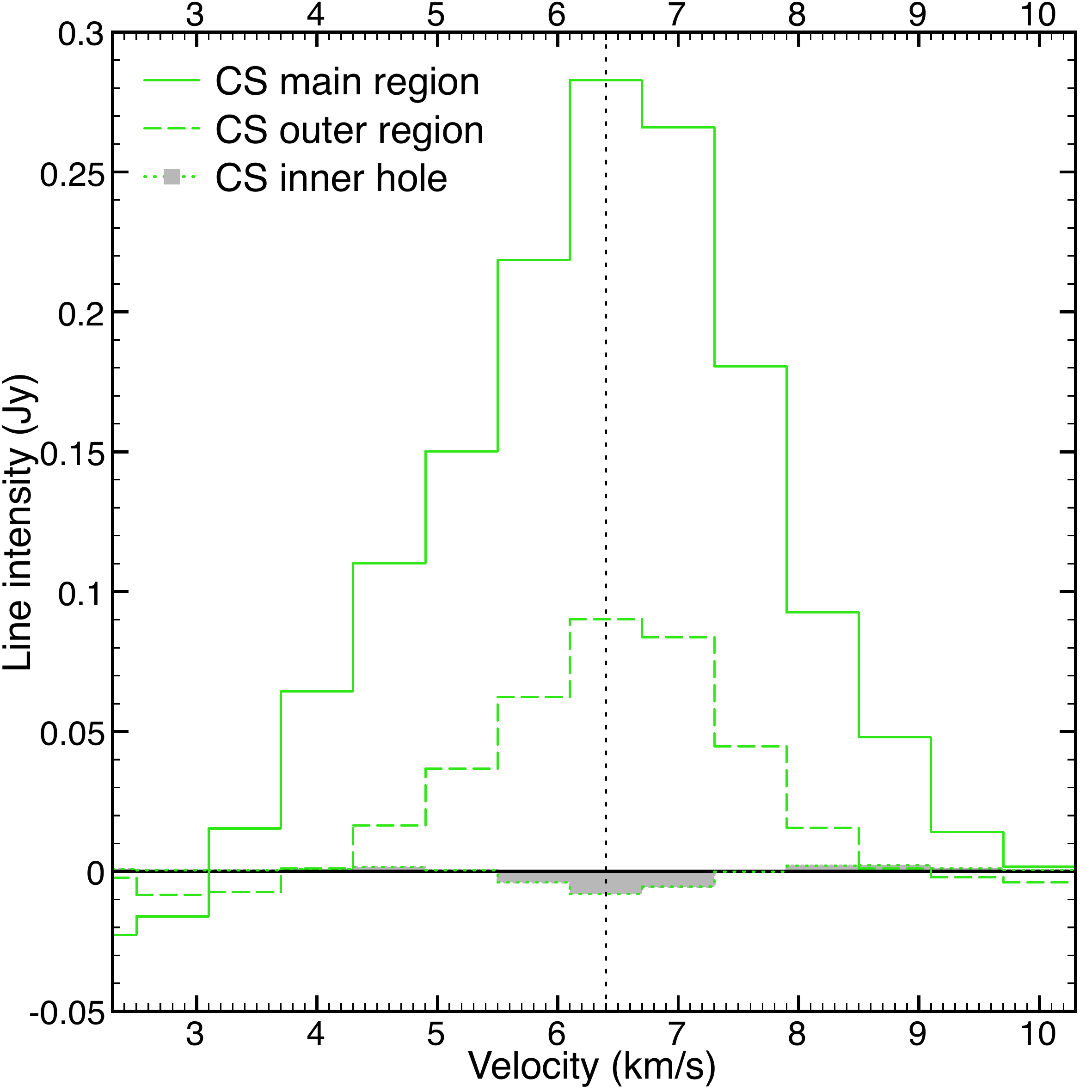}
   \includegraphics[width=6cm]{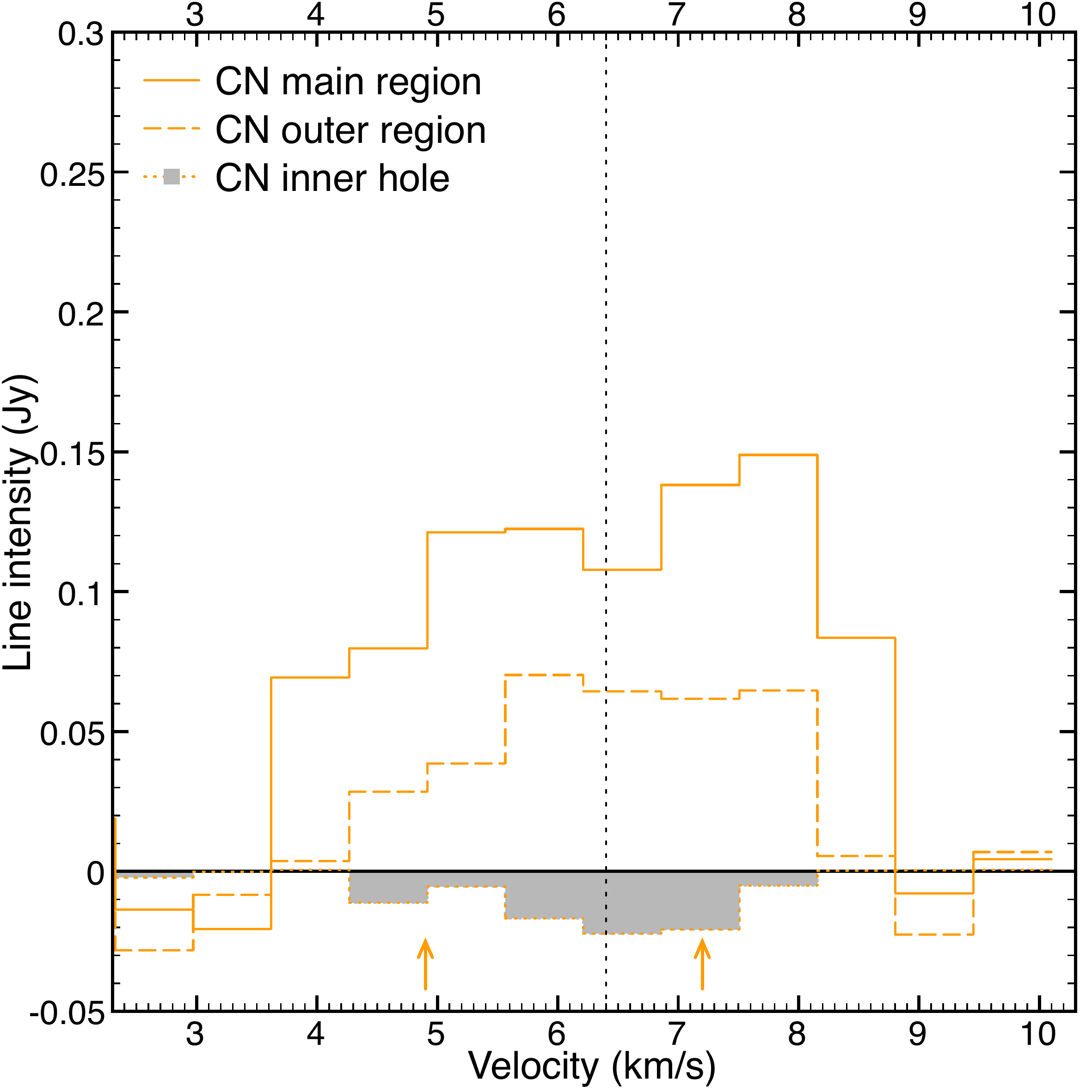}
    \includegraphics[width=6cm]{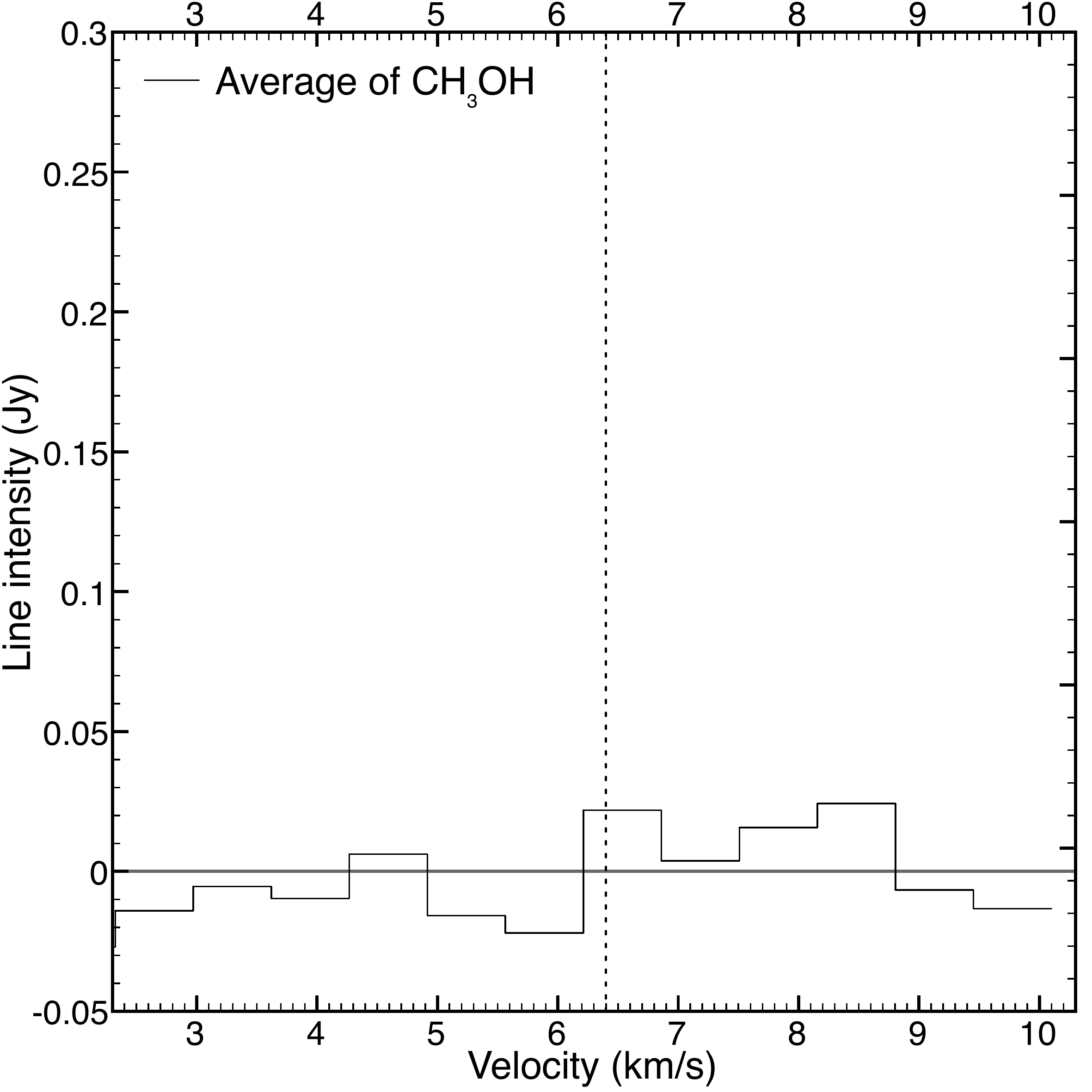}
   \includegraphics[width=6cm]{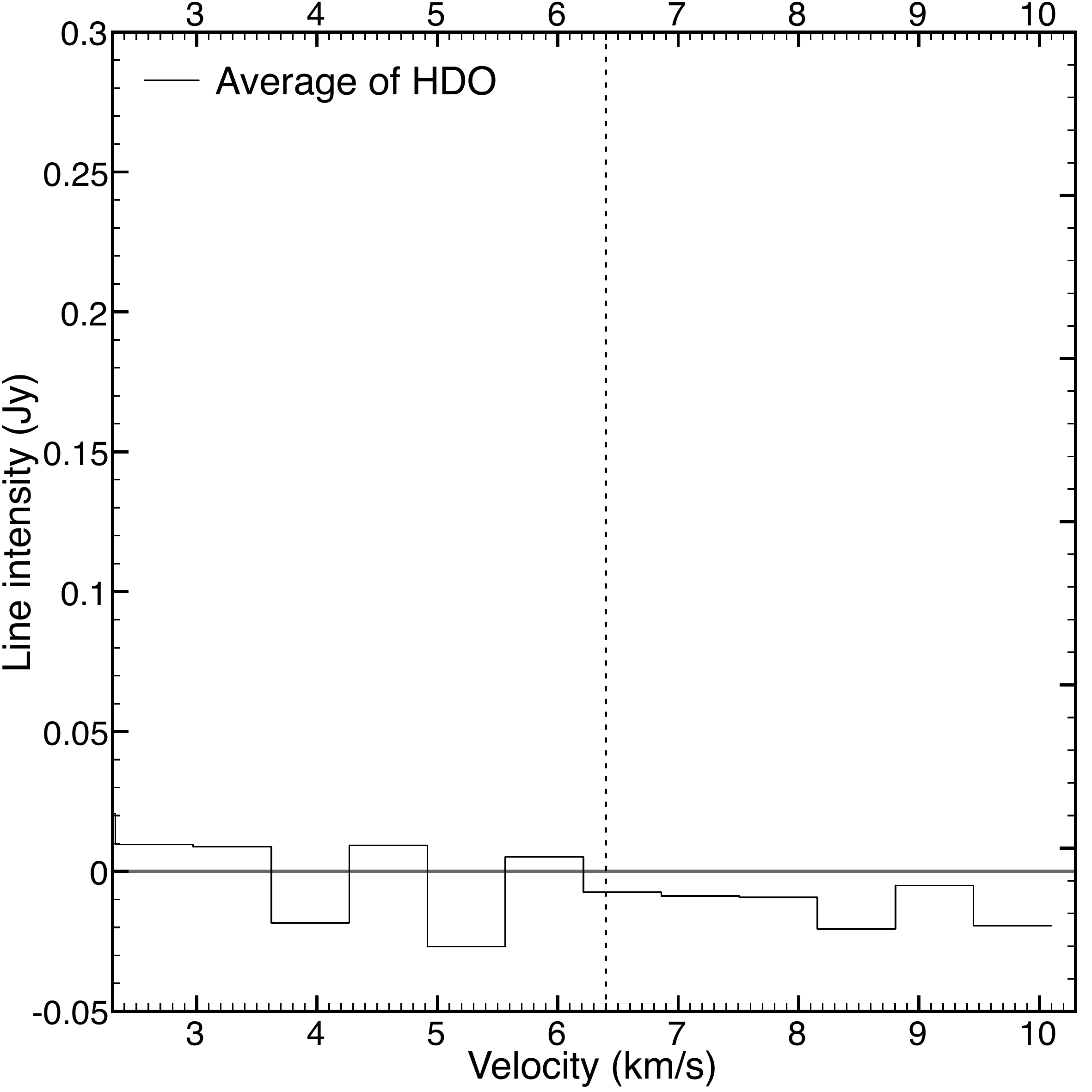}
    \includegraphics[width=6cm]{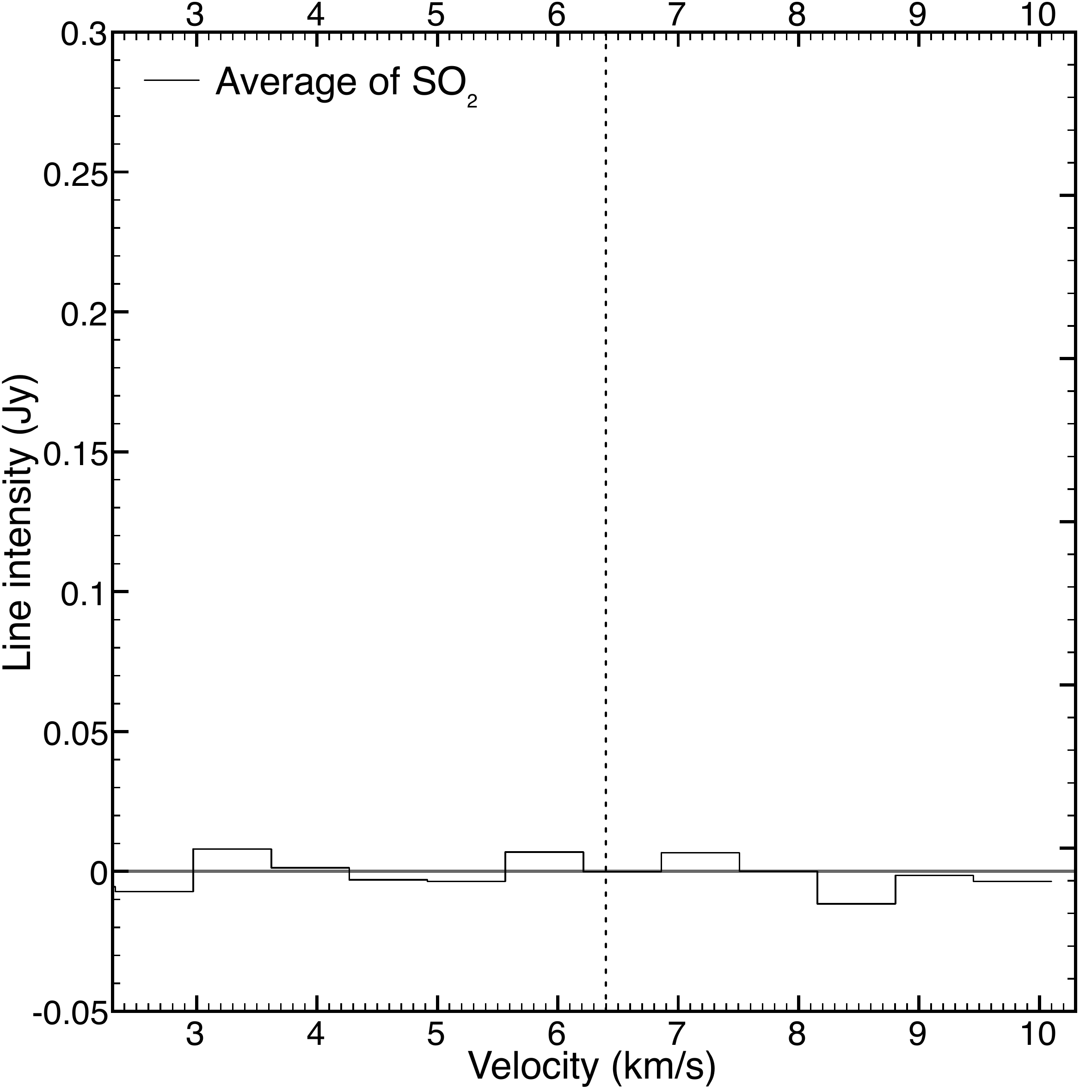}
      \caption{Spectral line profiles. Lines in the first row (detected) are integrated over three rings at different radii: 0\arcsec\ $-$ 0.2\arcsec\ (the inner hole), 0.5\arcsec\ $-$ 2.0\arcsec\ (the main region), and 3.5\arcsec\ $-$ 4.5\arcsec\ (the outer region). The shaded region highlights the profile from the inner hole. Arrows on the CN line profile denote the presence of blended lines. For lines in the second row (undetected) only the profile over the main region is shown. For these lines, the profile is obtained by {averaging} all available lines for that species. The vertical line indicates the systemic velocity of the source (6.4 km s$^{-1})$.}
          \label{Line_profile}
 \end{figure*}

Individual channel maps of CO, H$_2$CO, CS, and CN are shown in Figs.\,\ref{Channel_maps_1} and \ref{Channel_maps_2}. The spectral profiles obtained by integrating the emission at different disk regions are shown in Fig.\,\ref{Line_profile}. These regions are the inner 0.2\arcsec\ (inner hole), 0.5\arcsec$-$2.0\arcsec\ (main region), and 3.5\arcsec$-$4.5\arcsec\ (outer region).

\begin{figure*}
  \centering
 \includegraphics[width=18cm]{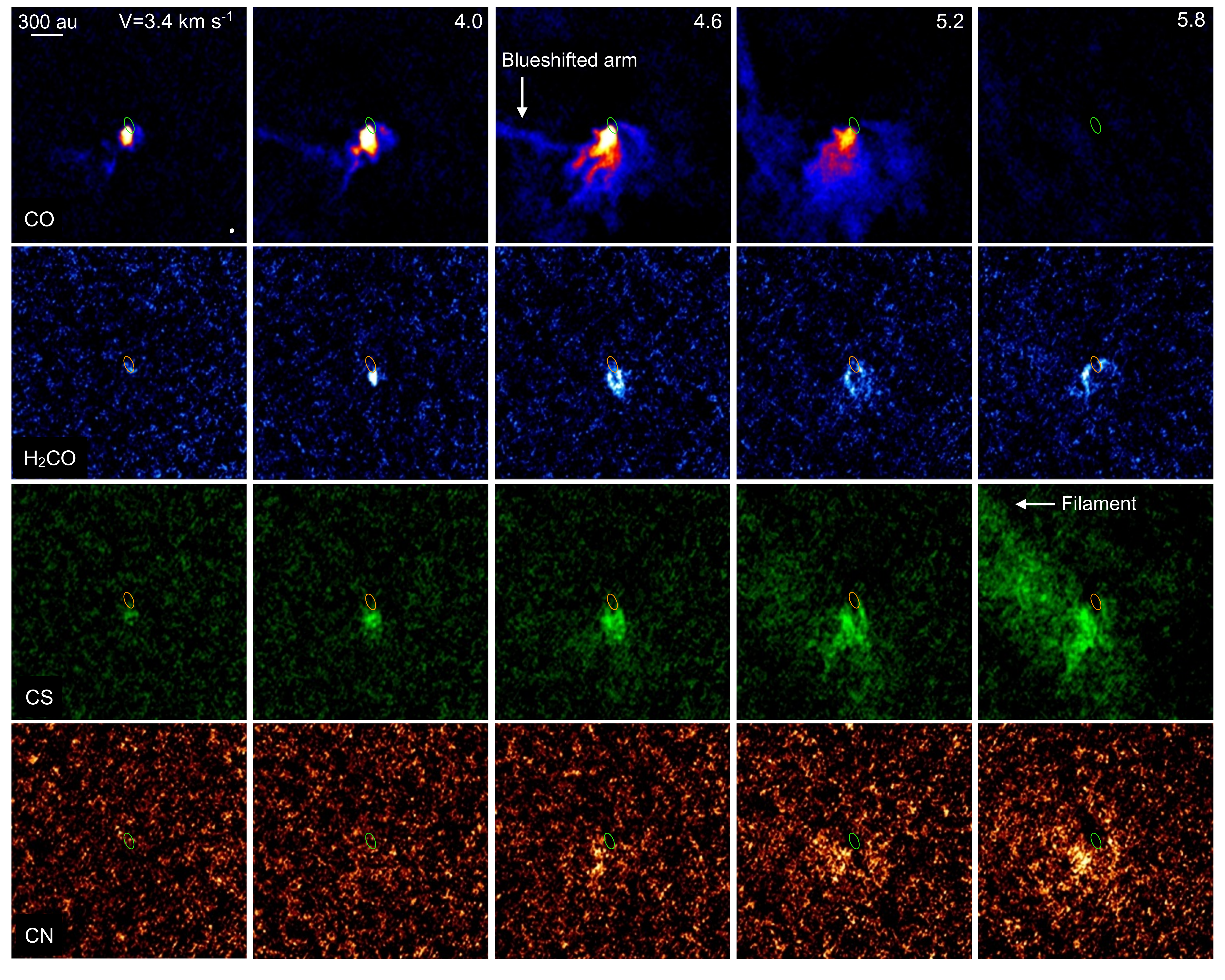}
     \caption{Channel maps of the molecular emission. The H$_2$CO line is shown along the first and fourth rows, the CS along the second and the fifth rows, and the CN along the third and the sixth rows. The inner ellipse indicate the innermost contour of the continuum emission from Fig.\,\ref{Continuum} (100$\sigma$ significance). {The beam size of all images is shown in the bottom left corner of the first panel.} North is up, east is left. It continues in Fig.\,\ref{Channel_maps_2}.} 
 \label{Channel_maps_1}
 \end{figure*}

\begin{figure*}
  \centering
 \includegraphics[width=18cm]{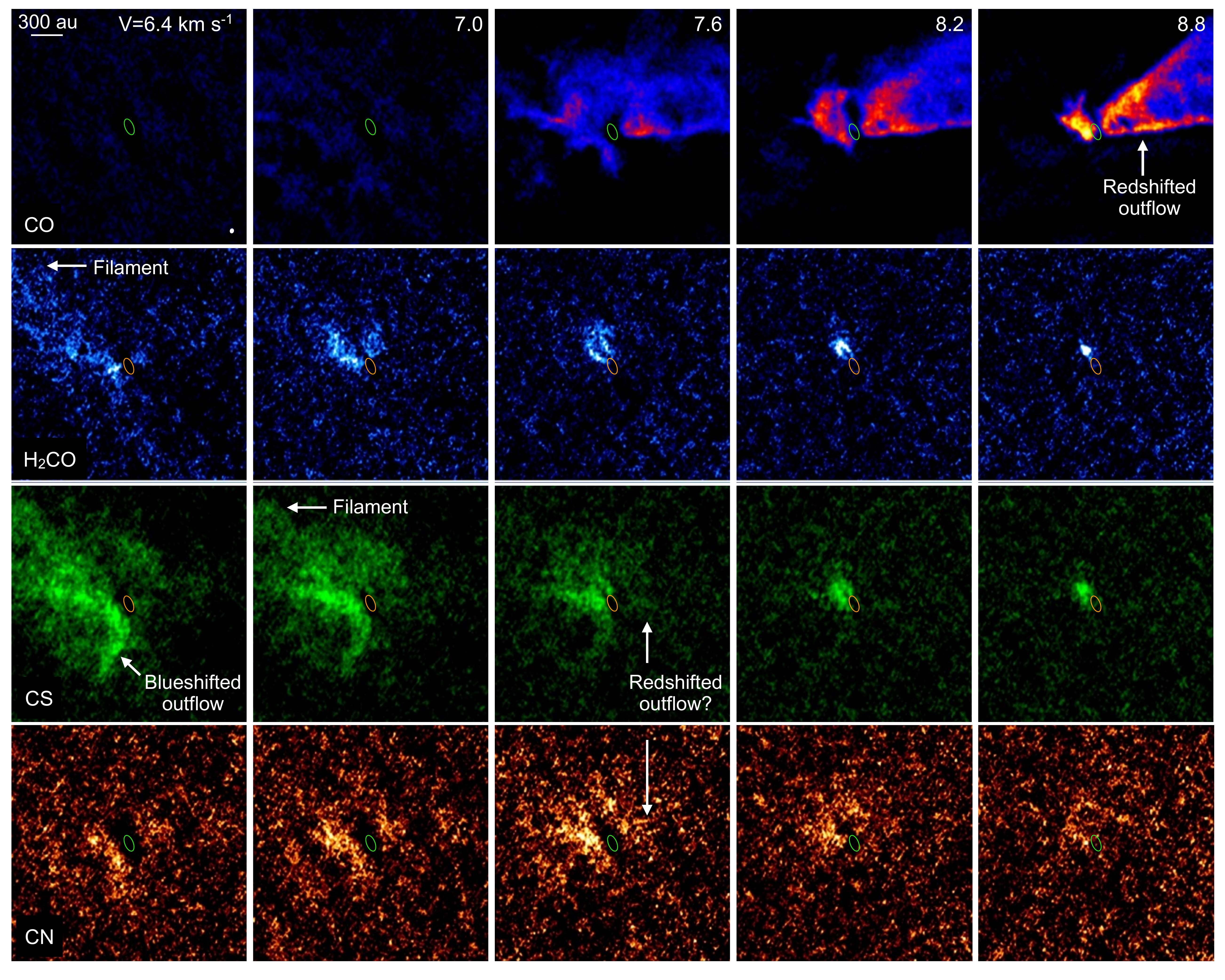}
     \caption{Continued from Fig.\,\ref{Channel_maps_1}.} 
 \label{Channel_maps_2}
 \end{figure*}

\end{appendix}

\end{document}